\documentclass[12pt,preprint]{aastex}

\newcommand{\ax}[1]{{1AXG~J#1}}
\newcommand{\ergs}{ergs s$^{-1}$ cm$^{-2}$}
\newcommand{\erg}{ergs s$^{-1}$}

\newcommand{\oiii}{{[O~III]$\lambda\lambda$5007,4959}}
\newcommand{\oiiia}{{[O~III]$\lambda$5007}}

\newcommand{\niib}{{[N~II]$\lambda$6583}}
\newcommand{\oii}{{[O~II]$\lambda$3727}}
\newcommand{\ciii}{{C~III]$\lambda$1909}}
\newcommand{\civ}{{C~IV$\lambda$1549}}
\newcommand{\mgii}{{Mg~II$\lambda$2800}}
\newcommand{\ha}{{H$\alpha$}}
\newcommand{\lya}{{Ly$\alpha$}}
\newcommand{\hb}{{H$\beta$}}
\newcommand{\apppi}{{$\Gamma_{\rm app}$}}

\slugcomment{ApJ Accepted, 2003 April 21}

\shorttitle{Optical Identification of {\it ASCA} MSSn}
\shortauthors{Akiyama et al.}

\begin{document}

\title{Optical Identification of the {\it ASCA}
Medium Sensitivity Survey in the Northern Sky :
Nature of Hard X-ray Selected Luminous AGNs\footnote{
Based on data collected at 8.2m Subaru Telescope,
which is operated by the National Astronomical Observatory of Japan,
University of Hawaii 2.2m telescope, Kitt Peak National Observatory
2.1m telescope, and Calar Alto 3.5m telescope.}}

\author{Masayuki Akiyama}
\affil{Subaru Telescope, National Astronomical Observatory of Japan,
Hilo, HI, 96720}
\email{akiyama@subaru.naoj.org}

\author{Yoshihiro Ueda}
\affil{Institute of Space and Astronautical Science, Sagamihara,
Kanagawa, 229-8510, Japan}
\email{ueda@astro.isas.ac.jp}

\author{Kouji Ohta}
\affil{Department of Astronomy,
Kyoto University, Kyoto, 606-8502, Japan}
\email{ohta@kusastro.kyoto-u.ac.jp}

\author{Tadayuki Takahashi}
\affil{Institute of Space and Astronautical Science, Sagamihara,
Kanagawa, 229-8510, Japan}
\email{takahasi@astro.isas.ac.jp}

\and

\author{Toru Yamada}
\affil{National Astronomical Observatory of Japan, Mitaka, 181-8588, Japan}
\email{yamada@optik.mtk.nao.ac.jp}

\begin{abstract}
We present the results of optical spectroscopic 
identifications of a bright subsample of 2--10~keV hard 
X-ray selected sources from the {\it ASCA} 
Medium Sensitivity Survey in the northern sky (AMSSn). 
The flux limit of the subsample is 
$3\times10^{-13}$ \ergs\ in the 2--10~keV band. 
All but one of the 87 hard X-ray selected sources are 
optically identified, with AGNs (including broad-line AGNs, 
narrow-line AGNs, and 3 BL Lac objects), 7 clusters 
of galaxies, and 1 galactic star. It is the largest 
complete sample of hard X-ray selected AGNs at the 
bright flux limit. Amounts of absorption to their nuclei
are estimated to be hydrogen column densities ($N_{\rm H}$) of 
up to $\sim3\times10^{23}$ cm$^{-2}$ from their X-ray spectra. 
Optical properties of X-ray absorbed AGNs with 
$N_{\rm H} > 1\times10^{22}$ cm$^{-2}$ indicate 
the effects of dust absorption: at redshifts, $z<0.6$,
AGNs without broad H$\beta$ emission lines have
significantly larger $N_{\rm H}$ value than 
AGNs with broad H$\beta$ emission lines. 
At $z>0.6$, the X-ray absorbed AGNs have a large hard 
X-ray to optical flux ratio ($\log f_{\rm 2-10~keV}/f_{R} > +1$).
However, three X-ray absorbed $z>0.6$ AGNs show strong broad lines.
In combination with hard X-ray selected AGN samples
from the {\it ASCA} Large Sky Survey, the {\it ASCA}
Deep Survey in the Lockman Hole and {\it Chandra} Deep Field North, 
the luminosity distributions of absorbed ($N_{\rm H} > 1\times10^{22}$
cm$^{-2}$) and less-absorbed ($N_{\rm H} < 1\times10^{22}$ cm$^{-2}$)
AGNs are compared. 

\end{abstract}

\section{INTRODUCTION}

Revealing the origin of the Cosmic X-ray Background (CXB)
is one of key issues in understanding the growth of central
massive black holes in galaxies \citep{bar01}. Recently, ultra deep
$>$1Ms pointing observations with {\it Chandra} resolve
80\%--90\% of the CXB into discrete sources in the 
0.5--10~keV energy range (Brandt et al. 2001; Rosati et al. 2002). 
Preliminary results of optical identifications of these sources 
suggest that they mostly originate 
from accretion processes in AGNs, although there are contributions 
from hot gas of elliptical galaxies and X-ray binaries of 
starburst galaxies at the faintest flux level \citep{hor01}.
Therefore the CXB is thought to be an integrated emission of
AGNs with various redshifts, luminosities, and types; 
in other words, it reflects the cosmic growing process of the 
central massive black holes in galaxies. 

A significant fraction of the accretion process 
is expected to be obscured by absorbing matter in AGNs (Comastri et al. 1995;
Gilli et al. 2001). 
The intensity and the spectrum of the CXB are well modeled if 
we assume that there are four times more absorbed 
AGNs than non-absorbed AGNs \citep{com95}.
In the model it is assumed that the fraction of
absorbed AGNs in low-luminosity AGNs, i.e. Seyfert galaxies,
can be simply extrapolated to high-luminosity AGNs, i.e. QSOs. In the local 
universe ($z<0.03$), the number density of narrow-line
low-luminosity AGNs (Seyfert 2s) is three times 
larger than that of Seyfert 1s (Maiolino \& Rieke 1995). 
On the other hand, 
for high-luminosity QSOs, only several candidates of absorbed 
narrow-line luminous QSOs have been found in contrast to the fact that 
several thousands of broad-line QSOs have been cataloged.
The model of the CXB expects that there are many 
absorbed luminous QSOs, which escape from traditional 
QSO survey methods with optical/UV or soft X-ray 
selections. However, there is another possibility that the 
number ratio between absorbed and non-absorbed AGNs
depends on intrinsic luminosities of AGNs and it decreases
with increasing luminosity. For this case,
the CXB can be dominated by absorbed 
Seyfert galaxies \citep{fra02}.
In order to specify the site (QSOs or Seyferts ?) of the major
growth of the central massive black holes of galaxies, 
it is very important to reveal the number density
of absorbed QSOs.

Since hard X-rays can penetrate absorbing materials of AGNs,
AGNs with hydrogen column density, $N_{\rm H}$, as large 
as $1\times10^{23}$ cm$^{-2}$ can be detected without a 
significant bias, using 2--10~keV hard X-ray emission.
Several candidates of high-redshift absorbed 
QSOs are found in optical spectroscopic identifications of 
hard X-ray selected sources 
(e.g. \citet{oht96} with {\it ASCA}, and \citet{ste02} 
with {\it Chandra}). Additionally, one third of the optical counterparts 
of the hard X-ray selected sources in the ultra deep {\it Chandra} survey have 
a faint optical magnitude, and most of the faint optical 
counterparts could be obscured QSOs \citep{ale01}. 
These discoveries of candidates of obscured QSOs imply that
a significant fraction of the QSO population has been missed because
of absorption to the nucleus. The fraction of absorbed QSOs 
in the whole QSO population is not clear.
The optical spectroscopic identifications of the
faint hard X-ray selected sources are not complete and insufficient to 
examine the fraction of absorbed QSOs in detail.
Nuclear emissions of absorbed QSOs in the optical 
wave-band are expected to be fainter than those of non-absorbed QSOs
due to dust absorption to their nuclei. Since absorbed
QSOs are difficult to identify in comparison with non-absorbed QSOs,
the limit in magnitude for optical identifications could introduce
bias against absorbed QSOs. Therefore, a sample of hard
X-ray selected sources with highly complete optical 
identifications is necessary to estimate the fraction of absorbed
QSOs.

In order to construct a sample of hard X-ray selected AGNs,
we performed optical identification of 34 hard X-ray-selected 
sources from the {\it ASCA} Large Sky Survey (hereafter ALSS; 
Ueda et al. 1999a; Akiyama et al. 2000 [AOY00]) which covers 5.4 
square degrees with a flux limit of $1\times10^{-13}$ \ergs\ 
($2-10$ keV) with the {\it ASCA} Solid-state Imaging Spectrometer 
(SIS; Burke et al. 1991). All but one source are identified with 
30 AGNs, 2 clusters of galaxies and 1 galactic star.\footnote{Later 
{\it Chandra} follow-up observation suggests that the source 
without an identification (AX~J131832+3259) is a fake source 
(Ueda et al. in preparation). It is consistent with the estimated 
number of fake sources in the ALSS sample.}
In the identified AGNs in ALSS, there is no high-redshift luminous
cousin of Seyfert 2 galaxies. The redshift distribution of 
narrow-line AGNs with $N_{\rm H} > 1\times10^{22}$ cm$^{-2}$ is
limited to $z<0.5$, in contrast to the existence of 15 
broad-line AGNs at $z>0.5$. The difference of the redshift 
distributions suggests a deficiency of narrow-line AGNs with
$N_{\rm H} > 1\times10^{22}$ cm$^{-2}$ 
and large intrinsic hard X-ray luminosity, 
$L_{\rm 2-10keV} >1\times10^{44}$ erg s$^{-1}$. 
A part of the high-redshift broad-line AGNs are, perhaps,
significantly absorbed only in the X-ray band.
However, the number of the hard X-ray selected AGNs is too small,
especially for luminous AGNs, to make a definitive conclusion.

To expand the sample of hard X-ray selected AGNs, 
we conducted an optical identification program of a bright subsample 
of hard X-ray selected sources in the {\it ASCA}
Medium Sensitivity Survey (AMSS; Ueda et al. 1999b; 
Ueda et al. 2001 [Paper I]). 
AMSS is a serendipitous source survey based 
on the Gas Imaging Spectrometer (GIS; Ohashi et al. 1996) 
data of {\it ASCA} pointing observations at high Galactic 
latitude region. From the catalog, 87 bright hard X-ray selected 
sources in the northern sky are selected above flux limit of 
$3\times10^{-13}$ \ergs\ ($2-10$ keV) for
the optical identification (hereafter AMSSn sample).
By intensive spectroscopic observations, all but one of the 87 
sources have been optically identified. 

The flux limits of the ALSS and AMSSn are 
2 orders of magnitudes shallower than those of 
deep {\it Chandra} and {\it XMM-Newton} surveys. 
But, the total area of the {\it ASCA} surveys is about 
$70$ degree$^{2}$, and 3 orders of magnitude larger than deep 
{\it Chandra} surveys. Thus, the {\it ASCA} AGNs cover 
a different region of the redshift versus luminosity diagram 
in comparison with the AGNs from the deep surveys, 
and the {\it ASCA} AGNs are more suitable for studies 
of QSOs in the intermediate redshift ($z<1$) universe 
(see Section 3.6). Additionally, the brighter hard X-ray 
selected sources of the {\it ASCA} surveys with brighter 
optical counterparts than the deep surveys make it 
possible to achieve a complete optical identification of 
hard X-ray selected sources. The {\it ASCA} AGNs provide 
us a unique opportunity to examine the fraction of 
absorbed QSOs in the intermediate redshift universe.
The model of the CXB predicts that 40\% of the AGNs detected
above the flux limit of {\it ASCA} should be significantly
absorbed, and the {\it ASCA} flux limit is sufficient to detect the 
tip of the iceberg of the absorbed AGN population which contributes 
to the CXB \citep{com95}. 

Details of the AMSSn sample are 
described in Section 2. The identifications are 
summarized in Section 3. From the identified 
hard X-ray selected sources, natures of hard X-ray selected 
AGNs are discussed and the fraction of absorbed AGNs in 
combination with the hard X-ray selected AGNs from ALSS, {\it ASCA}
Deep Survey in the Lockman Hole, and {\it Chandra} Deep Field
North are examined (Section 4). 
Radio properties of the AMSSn AGNs are discussed in Section 5.
Summary is given in Section 6. The hard X-ray
luminosity function of the hard X-ray selected AGNs is
discussed in a separate paper (Ueda et al. 2003).
In this paper, $H_{0}$ of 50 km s$^{-1}$ Mpc$^{-1}$ and
$q_{0}$ of 0.5 are used throughout.

\section{SAMPLE}

AMSS is a serendipitous source survey based on the 368 combined fields 
of {\it ASCA} GIS pointing observations at high Galactic 
latitude ($|b| > 10^{\circ}$) observed in the period
between 1993 May and 1996 December. 
In total, 370 sources are extracted from the {\it ASCA} GIS
data above 5$\sigma$ in the 2--10~keV band serendipitously.
In order to concentrate on bright hard X-ray selected
sources, for which a position and an X-ray spectrum are determined
relatively well, X-ray sources are selected with the criteria:
1) the detection significance in the 2--10~keV band is $> 5.5 \sigma$;
2) the Galactic-absorption-corrected countrate in the 2--10~keV band is 
$> 2.7$ counts ks$^{-1}$, which corresponds to 
$3 \times 10^{-13}$ \ergs\ if we assume an X-ray source with a power-law 
spectrum with a photon index of 1.7 and no absorption;
3) the distance from the center of GIS field of view is
$< 20^{\prime}$;
4) the Galactic latitude is $|b| > 30^{\circ}$; 
5) the declination is $> -20^{\circ}$;
and 
6) the source is not a primary target of the {\it ASCA} 
observation and is not physically related to the primary target.
With these criteria, 87 X-ray sources are selected in the hard
X-ray band from the original AMSS catalog. They
are listed in Table~\ref{id_table} with the significance
in the hard X-ray band ($\sigma$) and the X-ray coordinate. 
For convenience, each X-ray source is referred to as
its exact name and identification number, like
\ax{000605$+$2031(NE01)}. Galactic-absorption-corrected count rates 
of the sources in the hard and soft X-ray bands are listed 
in the ``Count rate'' column in Table~\ref{id_flux}. 
The Galactic hydrogen column density estimated
from HI observations \citep{dic90} is used (``$N_{\rm H}$'' 
in Table~\ref{id_flux}). 
The total survey area depends on the count rate limit: 
for example, 34 degree$^{2}$ and 
68 degree$^{2}$ with the count rate limits of 
2.7 counts ks$^{-1}$ and 10 counts ks$^{-1}$
in the 2--10~keV band, respectively
(for the survey area as a function of count rate limit, 
see Ueda et al. 2003). The former and the latter count rate
limits correspond to $3\times10^{-13}$ \ergs\ and $1\times10^{-12}$ \ergs\
for a power-law spectrum with a photon index of 1.7 and 
no absorption, respectively.

A difficulty for the optical identification of the 
X-ray sources detected by {\it ASCA}
is their positional uncertainties which are caused by 
(1) an error of the absolute satellite attitude determination;
(2) a combined error caused by source confusion and a
statistical fluctuation due to a limited number of photons;
and (3) other systematic errors unique to the GIS instruments,
such as the position linearization maps, 
the grid support structure, etc (Paper I).
Therefore, for the current identification project,
we select X-ray sources which have sufficient counts ($>5.5\sigma$),
such that, at least, the positional uncertainty caused by
(2) is suppressed as much as possible. 
For example, for a source with a high significance ($10\sigma$) 
within $15^{\prime}$ from the GIS field center, 
the positional uncertainty, i.e., 90\% error radius,  
is $0.\!^{\prime}94$, and
for a source with the lowest sigma ($5.5\sigma$) at $20^{\prime}$
from the GIS center, 
the total positional uncertainty goes up to $1.\!^{\prime}58$.
The positional uncertainty of each X-ray source is listed in
the ``Unct.'' column of Table~\ref{id_table}.

For each X-ray source, 
hardness ratio, $HR \equiv (H-S)/(H+S)$ where
$H$ and $S$ represent the countrates corrected for
Galactic absorption in the 2--10~keV and 0.7--2~keV bands respectively
(columns ``H'' and ``S'' in Table~\ref{id_flux}),
is available in Paper I.
The hardness ratio of the source can be converted to 
the {\it apparent} photon index (\apppi) of the best-fit 
power-law model using the response of the GIS instrument.
In this paper, the \apppi\ is referred to describe the apparent X-ray 
spectral property of each source in the 0.7--10~keV band.
The determined \apppi\ index is listed in 
Table~\ref{id_flux}.

\section{OPTICAL IDENTIFICATION}

\subsection{Selection of Candidates of Optical Counterparts of X-ray Sources}

In order to reveal the nature of the hard
X-ray sources, it is crucial to perform
optical spectroscopic observations of
optical counterparts. 
We selected candidates of optical counterparts
of the hard X-ray sources, using databases
of extragalactic objects, soft X-ray sources,
and radio sources.

Half of the AMSSn hard X-ray selected sources (37/87) 
have a cataloged AGN or a cluster of galaxies 
with a redshift within the error circle 
(details of literature information from NASA/IPAC 
Extragalactic Database (NED) are summarized in Section 3.3).

For the remaining 50 sources, the 
Automatic Plate Measuring machine (APM) catalog, which has 
limiting magnitude of $R$ of 21 mag \citep{mcm92},
was used to search for optical counterparts. 
The catalog is obtained from scans of 
glass copies of the Palomar Observatory Sky Survey (POSS) plates.
From the ALSS results, it is expected that 
most of the optical counterparts of the X-ray sources with
a hard X-ray flux $> 3\times10^{-13}$ \ergs\ are 
brighter than 20 mag in $R$-band (Figure 3 of AOY00).
The optical counterparts of most of the 
AMSSn X-ray sources should be detected in the APM catalog. 
The $3^{\prime}\times3^{\prime}$ 
optical images of the AMSSn source positions are shown in
left panels of Figure~1. The images are taken from 
the Digitized POSS plates, acquisition images 
during spectroscopic observation mentioned below, or
$R$-band images taken with the University of Hawaii 
88$^{\prime\prime}$ telescope. All images are centered on 
the AMSSn source positions, except for \ax{000605$+$2031(NE01)}, 
for which the center of the image is shifted.
Error circles of the AMSSn sources are indicated with 
a circle centered on the source position. 
Most of the AMSSn sources have more than 1 APM object in their 
error circles.

In order to select the most plausible optical counterpart of 
each AMSSn X-ray source, {\it ROSAT} serendipitous source 
catalogs; the first {\it ROSAT} source catalog of pointed 
observations with the high resolution imager (HRI), 
the second {\it ROSAT} source catalog of pointed observations 
with the position sensitive proportional counter (PSPC), 
and {\it ROSAT} PSPC WGA catalog \citep{whi94}, from High 
Energy Astrophysics Science Archive Research Center (HEASARC) 
are used.  Typical positional uncertainties 
are 10$^{\prime\prime}$ for the HRI sources 
and 30$^{\prime\prime}$ for the PSPC sources. 
AMSSn sources which do not
have a very hard X-ray spectrum can be detected in 
the soft X-ray {\it ROSAT} surveys. For example, 
an X-ray source with a 2--10~keV 
flux of the AMSSn survey limit, $3\times10^{-13}$ \ergs, 
the 0.5--2~keV flux is expected to be $1.3\times10^{-13}$ \ergs\
if the photon index of the source is 1.7 without absorption. 

Thirty three AMSSn sources, including sources with 
a cataloged AGN or cluster of galaxies, 
have a {\it ROSAT} HRI source within the error radius.
The positions of the HRI sources are marked in 
the left panels of Figure~1 with a $10^{\prime\prime}$ 
radius circle. Most of the HRI sources have one candidate 
optical counterpart. The candidate optical 
counterparts are listed in Table~\ref{id_table} with an indication
``RH'' in the selection column. In the HRI error circles of
two other sources, \ax{160118$+$0844(NO53)} and 
\ax{210738$-$0512(NO17)}, there is no optical object cataloged in 
the APM catalog. In both of the error circles, a faint optical object
is detected in a deeper image taken during spectroscopic 
observation. Thus the faint objects are picked up as candidates 
of the optical counterparts. In the HRI error circles of other two 
sources (\ax{131112$+$3228(NE22)} and \ax{164045$+$8233(NE07)}), 
there is no optical object above $R$ of 21 mag. They are already 
cataloged as clusters of galaxies, and X-ray spectra of the AMSSn 
sources are soft (\apppi\ $>2$). The HRI positions may 
correspond to a centroid of the whole cluster
emission from hot gas, thus the positions do not 
match any particular galaxy. 
 
Additionally, 48 AMSSn sources have {\it ROSAT} PSPC 
serendipitous sources within their error circles.
The positions of PSPC sources are marked in 
the left panel of Figure~1 with a $30^{\prime\prime}$ 
radius circle. 
APM objects in the PSPC error circles are observed
with high priority in spectroscopic observations.
The positions of sources from the {\it ROSAT} All-Sky Survey
(RASS; Voges et al. 1999) are also  
marked with a $45^{\prime\prime}$ radius circle
in the left panel.

For four AMSSn X-ray sources,
\ax{010952$-$1252(SE33)},\\ \ax{111432$+$4055(NE26)}, 
\ax{170730$+$2353(NO24)}, and\\ \ax{233200$+$1945(NO18)}, 
follow-up hard X-ray observations with {\it Chandra} are conducted 
(Ueda et al. in preparation), and 6.78$\sigma$, 4.15$\sigma$, 2.95$\sigma$, and
3.80$\sigma$ sources respectively are detected in the {\it ASCA} error 
circles of the sources. The positions of the detected sources are, 
$01^{\rm h} 09^{\rm m} 50.\!^{\rm s}90, 
-12^{\circ} 53^{\prime} 22.\!^{\prime\prime}37$,
$11^{\rm h} 14^{\rm m} 31.\!^{\rm s}92, 
+40^{\circ} 56^{\prime} 14.\!^{\prime\prime}41$,
$17^{\rm h} 07^{\rm m} 32.\!^{\rm s}10, 
+23^{\circ} 53^{\prime} 41.\!^{\prime\prime}95$,
and 
$23^{\rm h} 31^{\rm m} 58.\!^{\rm s}90, 
+19^{\circ} 44^{\prime} 37.\!^{\prime\prime}68$.
At the positions of the {\it Chandra} sources 
of \ax{010952$-$1252(SE33)}, 
\ax{111432$+$4055(NE26)} and \ax{170730$+$2353(NO24)}, 
there is an optical object. For \ax{233200$+$1945(NO18)}, 
no APM object is in the {\it Chandra} error circle. 
A faint optical object discovered in an image 
taken during a spectroscopic observing run is picked
up as the optical counterpart.
The optical objects are marked with ``CH'' in the selection 
column of Table~\ref{id_table}, and the positions 
of {\it Chandra} sources are marked with triangles 
in the left panels of Figure~1.

For \ax{233200$+$1945(NO18)} and \ax{234725$+$0053(NE23)},
follow-up spectroscopic observations with {\it XMM-Newton} are conducted
(Ueda et al. in preparation).
The position of \ax{233200$+$1945(NO18)} which is 
determined by the {\it Chandra} observation is confirmed.
For \ax{234725$+$0053(NE23)}, three X-ray sources are
detected near the {\it ASCA} position. The brightest
source is more than three times brighter than the other
sources, and consistent with the {\it ASCA} flux. 
The position of the brightest source is  
$23^{\rm h} 47^{\rm m} 21^{\rm s}, 
+00^{\circ} 54^{\prime} 27^{\prime\prime}$.
The optical object is marked with ``XM'' in the selection
column of Table~\ref{id_table}, and the position
of the {\it XMM-Newton} source is marked with a triangle
in the finding chart.

Cross-correlation between the AMSSn X-ray sources
and radio sources in the NRAO VLA Sky Survey 
(NVSS; Condon et al. 1998) in the northern sky 
($\delta> -40^{\circ}$) is also examined.
NVSS covers the entire area of the AMSSn survey with a 
flux density limit of 2.3 mJy at 1.4 GHz. 
The positional uncertainty of an NVSS source is less than $1^{\prime\prime}$
for relatively bright point 
sources with a flux larger than 15mJy, and goes up to
$7^{\prime\prime}$ for the faintest point sources with a flux density of
2.3 mJy \citep{con98}.
Figure~\ref{NVSS_dis} shows the stacked surface number density
distribution of the NVSS radio sources centered on 
the 87 AMSSn hard X-ray selected sources. 
There is an excess of radio source number 
density. Within $1^{\prime}$ from 
the positions of the 87 AMSSn sources, there are 
34 NVSS radio sources detected. From the number density of 
radio sources radius between $1^{\prime}$ and $3^{\prime}$,
the expected number of contaminating radio sources 
which are not related to an X-ray source is estimated 
to be $5.1\pm0.8$ for the 34 radio sources. 
Therefore at least 85\% of the radio sources detected
within $1^{\prime}$ from an X-ray source position
are likely to be physically related to the X-ray source.
The radio object might be interacting
with the X-ray object, and the radio object itself 
might not be the origin of the X-ray source.
The optical counterparts of the radio sources are
targets of optical spectroscopy with high priority.
The selected objects are indicated with ``N'' in the selection 
column of Table~\ref{id_table} and the positions of the NVSS radio sources
are marked with large squares in the left panels of Figure~1. 
Faint Images of the Radio Sky at Twenty-centimeters 
catalog of radio sources (FIRST; Becker, White, \& Helfand 1995) 
is also used. The survey covers only a part of the AMSSn survey areas with
a flux limit of 1mJy (5$\sigma$).
The positional uncertainty of the FIRST sources is about
1$^{\prime\prime}$ for a point source.
The positions of the radio sources are indicated with
small squares in the left panels of Figure~1, and an optical 
counterpart of a FIRST source is marked with ``F'' in 
Table~\ref{id_table}. 

The candidates of optical 
counterparts selected by the above methods
are listed in Table~\ref{id_table}. In Figure~1, the listed objects 
are marked with numbers which indicate identification numbers in APM finding 
charts. For AMSSn sources which have no information from 
other catalogs, the brightest and the bluest object
in the error circles is observed with high priority. 

\subsection{Spectroscopic Observations}

According to the list of the candidates of optical counterparts,
we performed the spectroscopic observations for all sources
of our samples, except for candidates of clusters of galaxies. 
We also obtained optical spectra for sources whose type and
redshift were previously known, in order to examine the strength
of any emission lines.
Most of the spectroscopic observations were 
conducted with the University of Hawaii $88^{\prime\prime}$ 
telescope and KPNO 2.1m telescope. Calar Alto 3.5m telescope
and 8.2m Subaru telescope were used for others.

The UH$88^{\prime\prime}$ observations were made on 
2000 March 24, 25, 26, and 27, 2000 October 4, 5, and 6, 
and 2001 March 19, 20, and 21 with the Wide Field Grism
Spectrograph. A grating of 420 grooves mm$^{-1}$ with
blaze wavelength of 6400{\AA} was used. The spatial resolution was
$0.\!^{\prime\prime}35$ pixel$^{-1}$ and the typical image size
during the observations was $0.\!^{\prime\prime}8 \sim 1.\!^{\prime\prime}2$.
A slit width of $1.\!^{\prime\prime}2$ was used.
The spectral sampling was 3.75{\AA} pixel$^{-1}$.
The wavelength range from 4000{\AA} to 9000{\AA} was covered without
an order cut filter. Obtained spectra were affected by the 
second-order component above 8000\AA. The spectral resolution, 
which was measured by the HgAr lines in comparison frames and 
night-sky lines in the object frames, was 12{\AA} (FWHM).
For the flux calibration of the data, 
Feige 34 and BD+28 were observed for spring- and autumn-run, 
respectively. The same setup including slit width for the objects is 
used for the standard stars.
Imaging data of each X-ray source are taken without a 
filter for finding charts. 

The KPNO 2.1m observations were made on 2000 March 
8 and 9 and 2000 October 20, 21, and 22 with the 
Gold Camera Spectrograph. A grating (\#32) with 300 
grooves mm$^{-1}$ and a blaze wavelength of 6750\AA\ was used.
The spectral sampling was 2.47 {\AA} pixel$^{-1}$. The
wavelength range from 4000{\AA} to 8000{\AA} was covered with
an order cut filter for the wavelength range shorter than
4000{\AA} (GG400). The spectral resolution was measured to 
be 8 {\AA} (FWHM) from night-sky lines in the object frames. 
The spatial sampling was $0.\!^{\prime\prime}78$ pixel$^{-1}$.
The typical image size during the observation was $2^{\prime\prime}$.
The slit width was $2^{\prime\prime}$ for objects. 
Feige 34 and BD+28 were observed as standard stars in the
March and October runs, respectively. 
The slit width of $10^{\prime\prime}$ was used for the
standard stars to collect its whole light.
The other setups are the same for the objects.

Three sources (\ax{144109$+$3520(NO32)},
\ax{144301$+$5208(NO26)}, \\ and \ax{150430$+$4741(NO12)}
were observed with the 3.5m telescope at Calar Alto observatory 
on 1999 April 7 with the MOSCA instrument in a single-slit mode. 
The g250 grating which has 250 grooves mm$^{-1}$ and a blaze 
wavelength of 5700{\AA} was used. The spectral coverage ranged 
from 4000{\AA} to 8000{\AA}. In the configuration, the 
spectral sampling was 5.95{\AA} pixel$^{-1}$. A slit width of $1.\!^{\prime\prime}5$, 
which was the same as the FWHM of the image size, was used. 
The spectral resolution was measured to be 24{\AA} (FWHM) 
from widths of night sky lines in the object frames. 
The spatial sampling was $0.\!^{\prime\prime}32$
pixel$^{-1}$. HD84937 was observed as a standard star
with the same setup for the objects. 

Two faint objects (\ax{233200$+$1945(NO18)} 
and \ax{010952$-$1252(SE33)}), were observed on
the 8.2m Subaru telescope with the Faint Object Camera And Spectrograph 
(FOCAS; Kashikawa et al. 2002) on 2001 July 18.
A 300 grooves mm$^{-1}$ grating with a blaze wavelength of
5500{\AA} (300B) and an order cut filter below 4700{\AA} (SY47) were
used. The spectral sampling was 2.8{\AA} bin$^{-1}$ with 2 pixel binning.
The spatial sampling was $0.\!^{\prime\prime}3$ bin$^{-1}$ with
3 pixel binning. Wavelength range from 4700{\AA} to 9000{\AA} 
was covered. The slit width was $0.\!^{\prime\prime}6$ and
the image size during the observation was $0.\!^{\prime\prime}7$.
The spectral resolution was measured to be 11{\AA} (FWHM) from 
night sky lines in the object frames.
For the flux calibration, HZ44 was observed as a standard 
star with a $2^{\prime\prime}$ slit to collect its whole light.
 
In order to obtain an optical spectrum with a high
signal to noise ratio for the hardest X-ray source \ax{170730$+$2353(NO24)},
the optical counterpart was observed on the Subaru telescope 
with FOCAS on 2002 June 7. The slit width of 
$0.\!^{\prime\prime}8$ was used and the spatial sampling was 
$0.\!^{\prime\prime}4$ bin$^{-1}$ with 4 pixel binning in the
spatial direction. All other settings were the same as the previous 
observation. The image size during the observation was
$0.\!^{\prime\prime}6$. The spectral resolution was measured to be 
9.5{\AA} (FWHM) from night sky lines in the object frames. 
Although the slit width of the observation is larger than 
the previous observing run, the spectral resolution is higher, 
probably due to better focusing of spectrograph during the 
later observing run. For the flux calibration, HZ44 was
observed as a standard star with a $2^{\prime\prime}$ slit.

The exposure time for each object is indicated at the top of
each spectrum in Figure~1.

All of the data are analyzed using IRAF\footnote{
IRAF is distributed by the National Optical Astronomy Observatories,
which is operated by the Association of Universities for
Research in Astronomy, Inc. (AURA) under cooperative 
agreement with the National Science Foundation.}.
After bias subtraction, flat-fielding,
and wavelength calibration, the optimum extraction method by {\bf apextract}
package is used to extract one dimensional spectral data from the two
dimensional original data. For the UH data, flux calibrations
do not work well for wavelengths $>$ 8000{\AA}, because
the spectra of the objects, especially for blue AGNs,  
and the standard stars are seriously affected by the 
second order spectra. Using the $R$ magnitude of each object,
we estimated the uncertainty of the flux calibration
to be factor of 1.7.

\subsection{Notes on Individual Identification}

The results of the spectroscopic observations 
are summarized in ``Classification'' and ``Redshift'' 
columns of Table~\ref{id_table}. Notes on an individual 
identification are summarized below. 
The $10^{\prime\prime}\times10^{\prime\prime}$
close-up views and the optical spectra of the identified objects 
are shown in the middle and right panels of Figure~1, respectively. 
Apparent and absolute properties of the identified objects in the X-ray, 
optical, near-infrared, and radio wavelengths are summarized 
in Tables~\ref{id_flux} and \ref{id_lum}, respectively. 

{\it \ax{000605$+$2031(NE01)}} --- There is a $ROSAT$ PSPC
source just outside of the error circle of the AMSSn source.
There are 2 bright objects (16\_04 and 14\_05) in the PSPC 
error circle. 16\_04 is a broad-line AGN at $z=0.385$ 
(Giommi et al. 1991; EXO0430.5-1252). 14\_05 is a K5-type star. 
If 16\_04 is an X-ray source, the optical-to-X-ray 
flux ratio, $\log f_{\rm 2-10~keV}/f_{\rm R} = -0.22$ is
in the range of other AGNs (Figure~\ref{opt_x1}).
If this X-ray source originates from a K-type star,
we expect that the star has $R$-band magnitude of 5.0 to 11.2 mag, 
based on the soft band flux of this source
($f_{\rm 0.3-3.5~keV} = 1.2\times10^{-12}$ 
erg s$^{-1}$ cm$^{-2}$ from soft band count rate and
\apppi),
typical optical-to-X-ray flux ratio of
K-type stars ($\log f_{\rm 0.3-3.5~keV}/f_{\rm V} = 
-4.0$ to $-1.5$ from \citet{sto91}), 
and $V-R$ of 0.9 mag \citep{pic98}.
The $R$-band magnitude of 14\_05 is fainter
than the expected magnitude range.
Therefore, 16\_04 is the most plausible optical 
counterpart of the X-ray source.   

{\it \ax{000927$-$0438(NO19)}} --- There is a $ROSAT$ PSPC
source detected in the error circle of the AMSSn source.
12\_01 is the brightest object in the PSPC error circle. 
It is identified with a broad-line AGN at $z=0.314$.

{\it \ax{001913$+$1556(NE18)}} --- There is a $ROSAT$ PSPC
source detected in the error circle of the AMSSn source.
There is an NVSS radio source detected at the edge of the
PSPC error circle. Optical counterpart of the NVSS source
(08\_02) is a broad-line AGN at $z=2.270$ (Marshall et al. 1983; ISS35). 

{\it \ax{002619$+$1050(NO49)}} --- There is a $ROSAT$ HRI
source in the error circle of the AMSSn source. 
Optical counterpart of the HRI source (30\_01) is
identified with a broad-line AGN at $z=0.474$.

{\it \ax{002637$+$1725(NO50)}} --- There are 2 bright galaxies
in the error circle of the X-ray source (19\_01 and 20\_03).
20\_03 is a broad-line AGN at $z=0.043$, and 19\_01 is a companion 
galaxy with no emission line. The redshift difference of the 
two galaxies corresponds to a velocity difference of $180$ km s$^{-1}$.
There is a faint $ROSAT$ HRI source at $1.\!^{\prime}3$ north
from the AMSSn source. 
With an X-ray spectrum of a power-law with the photon 
index of 1.7, the expected hard X-ray flux of the HRI 
source is well below the flux limit of the AMSSn sample,
and the $ROSAT$ HRI position is well outside of the uncertainty 
area of the AMSSn source, thus the HRI soft X-ray source is not
related to the AMSSn source. 

{\it \ax{010952$-$1252(SE33)}} --- This source is observed by
{\it Chandra}. The optical counterpart of the {\it Chandra} source (17\_00)
shows strong blue continuum with Ca H\&K and G-band absorption lines
at $z=0.505$, but without strong emission lines. 
The optical spectrum is similar to that of a BL Lac object. 
The object is detected in the NVSS survey, and
the radio flux is 3.2mJy. The estimated luminosity 
$L_{\rm 1.4GHz}$ is $3.5\times10^{24}$ W Hz$^{-1}$, 
and close to those of radio-loud AGNs in AMSSn sample
(see Section 5)

{\it \ax{015840$+$0347(NO46)}} --- The brightest object in
the error circle of the AMSSn source (17\_01) is
a broad-line AGN at $z=0.658$ (Lewis et al. 1979; UM153). 

{\it \ax{023520$-$0347(NO16)}} --- There is a $ROSAT$ PSPC
source in the error circle of the AMSSn source. 
The brightest object in the PSPC error circle (22\_01) is 
a broad-line AGN at $z=0.376$ (Stocke et al. 1991; MS0232.8$-$0400). 

{\it \ax{033516$-$1505(SE20)}} --- There are four optical objects
in the error circle of the AMSSn source (16\_00, 24\_00, 37\_02, 
and 33\_01). 37\_02 and 33\_01 are emission line objects.
The former is a narrow-line AGN at $z=0.122$ and the latter
is a narrow-line AGN at $z=0.501$. For the line ratios diagram,
see Figure~\ref{line_rat}. The AMSSn source is identified with 37\_02,
because there is a good correlation between hard X-ray 
flux and \oiiia\ line flux for AGNs \citep{mul94}, and the object has 5 times 
larger \oiiia\ flux than 33\_01 (Figure 1). 16\_00 is a G-type star, and
24\_00 is an object without a strong emission line.

{\it \ax{035008$-$1149(SE37)}} --- There is a $ROSAT$ PSPC 
source in the error circle of the AMSSn source. 
The brightest optical object in the PSPC error circle (15\_00)
is identified with a broad-line AGN at $z=0.459$.
An NVSS source is detected in the error circle of the 
AMSSn source, but there is no optical counterpart
in the error circle of the NVSS source.

{\it \ax{035137$-$1204(SE17)}} --- There is a RASS source
in the error circle of the AMSSn source. An optical object near 
to the center of the RASS source (22\_02) is identified with a 
broad-line AGN at $z=0.182$.

{\it \ax{041757$+$0101(NE25)}} --- There is a $ROSAT$ PSPC source
in the error circle of the AMSSn source.
There are two optical objects in the PSPC error circle, and
27\_01 is a G-type star and 27\_02 is a broad-line AGN 
at $z=0.126$. The G-type star is too faint to be the hard X-ray source,
and the AMSSn source is identified with the broad-line AGN.

{\it \ax{043420$-$0822(NE03)}} --- There is a $ROSAT$ HRI source
in the error circle of the AMSSn source. An optical object in 
the error circle (09\_01) is identified with a broad-line 
AGN at $z=0.154$. 

{\it \ax{044749$-$0629(NO08)}} --- The brightest object in the
AMSSn error circle (10\_02) is identified with a broad-line 
AGN at $z=0.213$.

{\it \ax{083747$+$6513(NO06)}} --- There is a
$ROSAT$ PSPC source detected in the error circle of the AMSSn source.
In the PSPC error circle, an NVSS radio source is detected.
The optical counterpart of the NVSS source (18\_01) is
a broad-line AGN at $z=1.105$ (Smith et al. 1976; 3C204). 

{\it \ax{090053$+$3856(NO54)}} --- There is an NVSS radio source
detected in the error circle of the AMSSn source.
The optical counterpart of the NVSS source (37\_01,
the brighter component at the east side) 
is a narrow-line AGN at $z=0.229$ (Allington-Smith et al. 1985; 0857+39). 

{\it \ax{090720$+$1639(NO36)}} --- There is a  
cluster of galaxies at $z=0.073$ (Stocke et al. 1991; Abell 0744)
in the error circle of the AMSSn source. $ROSAT$ sources
are detected in and around the cluster of galaxies.
The X-ray spectrum of the AMSSn source is soft 
(\apppi\ $=+2.51\pm0.07$), and is consistent 
with X-ray spectra of clusters of galaxies.

{\it \ax{102337$+$1936(NE12)}} --- There is a 
$ROSAT$ HRI source in the error circle of the AMSSn source.
The brightest object in the HRI error
circle (16\_01) is identified with a broad-line AGN at $z=0.407$.

{\it \ax{103934$+$5330(NO11)}} --- There is a 
$ROSAT$ HRI source in the error circle of the AMSSn source.
The brightest object in the HRI error
circle (12\_01) is identified with a broad-line AGN at $z=0.229$.

{\it \ax{104026$+$2046(NO41)}} --- There is a
FIRST radio source detected in the error circle of the AMSSn source.
An optical counterpart of the radio source (AA) is an elliptical 
galaxy at $z=0.240$ and does not show a strong emission line. 
A blue object ($B-R=0.51$) 
near the radio source (12\_01) is identified with a 
broad-line AGN at $z=0.467$. The AMSSn source is identified 
with the broad-line AGN.

{\it \ax{105722$-$0351(NE14)}} --- There are two $ROSAT$ 
HRI sources detected in the error circle of the AMSSn source. 
The southern one is nearer to the center
of the AMSSn source and is three times brighter than the other one.
An optical object (20\_02) in the error circle
of the southern HRI source is identified with a
broad-line AGN at $z=0.555$ (Stocke et al. 1991;
MS1054.8-0335). 

{\it \ax{111432$+$4055(NE26)}} --- This source is observed
by {\it Chandra}. The optical counterpart of the
{\it Chandra} source (29\_02) is
identified with a broad-line AGN at $z=0.153$.

{\it \ax{111518$+$4042(NO56)}} --- There is a $ROSAT$ HRI source
detected at the edge of the error circle of the AMSSn source.
The optical counterpart of the HRI source (19\_05) 
is a broad-line AGN at $z=0.079$ (Stocke et al. 1991; MS 1112.5$+$4059).

{\it \ax{121328$+$2938(NO28)}} --- There is a $ROSAT$ HRI source
detected in the error circle of the AMSSn source. In the error
circle of the HRI source, there are two optical objects,
24\_01 and 26\_02. 24\_01 is a 
broad-line AGN at $z=0.143$ (Appenzeller et al. 1998; RX~J1213.4+2938). 
This hard X-ray source is identified with the broad-line AGN.
26\_02 is a companion galaxy with HII-region like 
narrow emission lines (see Figure~\ref{line_rat}). 
The redshift difference of the two galaxies corresponds to
the velocity difference of $90$ km s$^{-1}$.

{\it \ax{121359$+$1404(NO01)}} --- There is a $ROSAT$ PSPC and 
a FIRST radio source (2.9mJy) detected in the error circle of the AMSSn 
source. The optical object (15\_02) at the center of the 
$ROSAT$ PSPC source is identified with a broad-line AGN at
$z=0.154$. The optical counterpart (20\_03) of the FIRST radio source
is an elliptical galaxy without strong emission line
at the same redshift. The redshift difference of the two galaxies
corresponds to the velocity difference of 540 km s$^{-1}$. The 
AMSSn source is identified with 15\_02.

{\it \ax{121427$+$2936(NO27)}} --- There is a $ROSAT$ HRI source
in the error circle of the AMSSn source. The optical object 
in the error circle of the HRI source (24\_01) is a
broad-line AGN at $z=0.309$ (Appenzeller et al. 1998; RX~J1214.4$+$2936). 

{\it \ax{121752$+$3006(NE02)}} --- There is a $ROSAT$ HRI source
in the error circle of the AMSSn source. The optical counterpart
of the HRI source (23\_01) is an AGN at $z=0.130$ (Strittmatter et al. 1972).
The object shows a strong continuum without
any strong emission lines, and is a BL Lac object.

{\it \ax{121854$+$2957(NO07)}} --- There is a $ROSAT$ PSPC source
in the error circle of the AMSSn source. The brightest object
in the PSPC error circle (17\_01) is a narrow-line AGN
at $z=0.178$. For the line ratio diagnostic of the object, 
see Figure~\ref{line_rat}.
This AMSSn source is identified with 17\_01.
It is a very red object ($J-K_{S} = 2.9$), 
and is discussed in Section 4.2.3.
08\_07 is an elliptical galaxy at the same redshift.
02\_00, which is located outside of the finding chart, is another 
companion galaxy of 17\_01 with HII-region like narrow-emission lines
(see Figure~\ref{line_rat}). The redshift difference between 17\_01 
and 02\_00 corresponds to the velocity difference of 1500 km s$^{-1}$.

(This source is also detected by $Beppo-SAX$, and identified
with an AGN at $z=0.18$ \citep{fio99}. The optical spectrum
shown in their paper is that of 17\_01, but, the coordinate of the
optical counterpart in their list is that of 08\_07.
The $J$- and $K_{S}$-bands magnitudes listed in \citet{mai00}
are also wrong.)

{\it \ax{121930$+$0643(NO44)}} --- There is a $ROSAT$ HRI source
detected in the error circle of the AMSSn source. 
The optical counterpart of the HRI source (17\_01) is 
a broad-line AGN at $z=0.081$ (Stocke et al. 1991; MS1217.0+0700).
The spectrum of the object shows strong FeII emission lines
around 5000\AA\ in observed frame, and
the X-ray spectrum is relatively soft (\apppi\ $=+2.22\pm0.10$). 
These properties are similar 
to narrow-line Seyfert 1s. But, 
the FWHMs of broad H$\alpha$ and H$\beta$ lines
($\sim$ 3500 km s$^{-1}$) are larger than
those of narrow-line Seyfert 1s ($<2000$ km s$^{-1}$).

{\it \ax{121930$+$7532(NO39)}} --- There is a $ROSAT$ HRI source
detected at the edge of the error circle of the AMSSn source.
The optical counterpart of the HRI source (15\_02) 
is a broad-line AGN at $z=0.464$ 
(Stocke et al. 1991; MS 1217.4$+$7549).

{\it \ax{122003$-$0025(NO03)}} --- The brightest optical object
in the error circle of the AMSSn source is 
a broad-line AGN at $z=0.422$ (Croom et al. 2001; 
2QZ J122004.3$-$002540).

{\it \ax{122017$+$0641(NO45)}} --- There is a $ROSAT$ HRI source
detected in the error circle of the AMSSn source. 
The optical counterpart of the HRI source (06\_01) is a
broad-line AGN at $z=0.287$.

{\it \ax{122049$+$7505(NO37)}} --- There is a $ROSAT$ HRI source
detected in the error circle of the AMSSn source.
The optical counterpart of the HRI source (21\_01) is 
a broad-line AGN at $z=0.650$
(Stocke et al. 1991; MS1218.7$+$7522).

{\it \ax{122135$+$7518(NO38)}} --- There is a $ROSAT$ HRI source
detected in the error circle of the AMSSn source.
The optical counterpart of the HRI source (24\_01) is
a broad-line AGN at $z=0.073$ (Arakelian et al. 1970; Mrk205). 

{\it \ax{122155$+$7525(NO40)}} --- There is  
a cluster of galaxies at $z=0.24$\\ (Stocke et al. 1991; MS1219.9$+$7542)
in the error circle of the AMSSn source. 
There is a $ROSAT$ HRI source detected just outside of the error circle 
of the AMSSn source. The optical counterpart of the HRI source
(24\_03) shows narrow-emission lines with AGN-like line ratios 
(see Figure~\ref{line_rat}). The X-ray spectrum of the AMSSn source 
is too hard (\apppi\ $=+1.69\pm0.23$) to be a cluster
of galaxies, thus the AMSSn source is identified with 
the narrow line AGN at $z=0.239$.

{\it \ax{122645$-$0037(NE05)}} --- There is an NVSS radio source
detected in the error circle of the AMSSn source. The optical
counterpart of the radio source is an elliptical galaxy at $z=0.16$.
Considering that the X-ray spectrum of the AMSSn source
is soft (\apppi\ $=+2.44\pm0.20$) 
and there is a weak excess of faint galaxies around the 
elliptical galaxy, the AMSSn source is identified with 
a cluster of galaxies at $z=0.16$.

{\it \ax{123605$+$2613(NE04)}} --- There are a $ROSAT$ PSPC source
and a FIRST radio source in the error circle of the AMSSn source.
The optical counterpart of the FIRST radio source (11\_00) shows strong
\oiii\ and \oii\ emission lines.
Its \oiiia\ to H$\beta$ flux ratio is large and consistent
with those of Seyfert 2 galaxies. 
The object is identified with an narrow-line AGN at $z=0.459$.

{\it \ax{125732$+$3543(NO31)}} --- There is a $ROSAT$ PSPC source
detected in the error circle of the AMSSn source. A blue object 
(15\_01; $B-R=0.36$) detected in the error circle of the PSPC 
source is a broad-line AGN at $z=0.524$ \citep{mar83}.

{\it \ax{125812$+$3519(NE16)}} --- There is a $ROSAT$ HRI source
detected in the error circle of the AMSSn source. 
The optical counterpart of the HRI source (12\_01) is identified
with a broad-line AGN at $z=0.310$. It is originally 
identified with an AGN at $z=2.04$ by \citet{wee85} (WEE83). 
MgII emission line at $z=0.310$
seems to be mis-identified with Ly$\alpha$.
A strong broad H$\alpha$ emission line may originate from uncertainty 
of flux calibration above 8000\AA.

{\it \ax{125828$+$3528(NE15)}} --- There is a $ROSAT$ PSPC
source detected in the error circle of the AMSSn source. 
A FIRST radio source is in the error circle of the PSPC source.
The optical counterpart of the
FIRST radio source (14\_02) is a broad-line AGN
at $z=1.900$ (Braccesi, Lynds, \& Sandage 1968; B194).

{\it \ax{130407$+$3533(NO30)}} --- There is no optical object
in the error circle of the AMSSn source. Just outside of the
error circle, a RASS source is detected. The brightest
object in the error circle of the RASS source is 
a broad-line AGN at $z=0.329$ (Marshall et al. 1983)

{\it \ax{130453$+$3548(NO29)}} --- There are two bright optical
objects (26\_02 and 22\_04) in the error circle of the AMSSn source.
26\_02 shows narrow emission lines with HII region like line ratio
at $z=0.034$ 
(see Figure~\ref{line_rat}). 22\_04 is a Galactic G-star.  
An optical object just outside of the
AMSSn error circle (20\_01) is
a broad-line AGN at $z=0.316$.
We identified this source with 20\_01.

{\it \ax{131112$+$3228(NE22)}} --- There is a cluster of galaxies
at $z=0.245$\\ (MS 1308.8$+$3244; Stocke et al. 1991) in the error circle
of the AMSSn source.
The X-ray spectrum of the source is soft (\apppi\ 
$=+2.32\pm0.22$), and consistent with 
that of a cluster of galaxies.
There are $ROSAT$ HRI and PSPC sources detected in 
the error circle of the AMSSn source.
There is no bright optical object in the $ROSAT$ HRI error circle.

{\it \ax{132310$-$1656(SE34)}} --- There is an NVSS source at
the edge of the error circle of the AMSSn source. The optical
counterpart of the NVSS source (21\_04) is a broad-line AGN 
at $z=0.022$. 45\_01 is a narrow-emission line galaxy with HII-region like
emission line ratios at $z=0.135$ (see Figure~\ref{line_rat}).
We identified this source with 21\_04.

{\it \ax{133937$+$2730(NE17)}} --- There is a $ROSAT$ HRI
source in the error circle of the AMSSn source. In the HRI
error circle, there are three optical objects (44\_00, 39\_02,
and 42\_00). 42\_00, at south-east in the HRI error circle, 
is a broad-line AGN at $z=0.908$.
39\_02 shows no emission line, and 44\_00 is an M-type star.
Thus, the AMSSn source is identified with the broad-line AGN.

{\it \ax{134412$+$0016(NE20)}} --- There is a $ROSAT$ PSPC
source in the error circle of the AMSSn source. In the PSPC
error circle, a FIRST radio source is detected. The optical
counterpart of the radio source (17\_00) is an emission line galaxy
at $z=0.452$. There is a hint of a broad-H$\beta$ emission
line, the X-ray source is identified with the AGN.

{\it \ax{134450$+$0005(NE19)}} --- There is a $ROSAT$ PSPC
source in the error circle of the AMSSn source. The brightest
object in the PSPC error circle (27\_01) is identified with a
broad-line AGN at $z=0.087$. 

{\it \ax{134741$-$1122(SE30)}} --- The brightest object in the
AMSSn error circle (19\_01)
is identified with a broad-line AGN at $z=0.100$.

{\it \ax{140528$+$2224(NO15)}} --- There is a $ROSAT$ HRI source
in the error circle of the AMSSn source. The optical counterpart
of the HRI source (25\_01) is a broad-line AGN at $z=0.156$
(Mason et al. 2000; RXJ140528.3$+$222331). 

{\it \ax{140532$+$5055(NO13)}} --- There are 5 bright objects
in and at the edge of the error circle of the AMSSn source.
24\_00 is identified with a broad-line AGN at $z=0.106$.
27\_03 and AA show elliptical galaxy-like continua, and 
are at the
same redshift as the broad-line AGN. 21\_01 and 29\_05 are
Galactic stars. We identified this source with 24\_00.

{\it \ax{141240$-$1209(SE03)}} --- There is a $ROSAT$ HRI source
detected at the edge of the error circle of the AMSSn source.
The optical counterpart of the HRI source (05\_03) is identified with
a broad-line AGN at $z=0.247$.

{\it \ax{141343$+$4340(NE27)}} ---  There is a cluster
of galaxies at $z=0.089$ (Crawford et al. 1995; Abell 1885).
The cD galaxy of the cluster, 35\_05, is a narrow-line AGN
\citep{cra95}. The X-ray spectrum of the AMSSn source is 
soft (\apppi\ $=+2.39\pm0.07$) and consistent 
with that of clusters of galaxies. 
Thus the AMSSn source is identified with the cluster of galaxies. 

{\it \ax{141426$-$1209(SE28)}} --- There is a $ROSAT$ HRI source
in the error circle of the AMSSn source. The optical counterpart
of the HRI source (14\_02) is identified with a broad-line AGN
at $z=1.156$.

{\it \ax{142353$+$2247(NE09)}} --- There is a $ROSAT$ HRI source
in the error circle of the AMSSn source. The optical counterpart
of the HRI source (57\_01) is identified with a narrow-line AGN
at $z=0.282$. For the line ratio diagnostic of the object,
see Figure~\ref{line_rat}.

{\it \ax{142651$+$2619(NO51)}} --- There is a $ROSAT$ PSPC source
in the error circle of the AMSSn source. The brightest object
in the PSPC error circle (35\_01) is identified with a broad-line AGN
at $z=0.258$. 
An NVSS source outside of the AMSSn error circle is
identified with a narrow-line AGN at $z=0.079$ (53\_07).
Because 35\_01 is detected by PSPC and is closer to the
center of the AMSSn source than 53\_07, the AMSSn source
is identified with 35\_01.

{\it \ax{144055$+$5204(NE13)}} --- There is a $ROSAT$ PSPC
source in the error circle of the AMSSn source. The brightest
object in the PSPC error circle (42\_02) is 
a broad-line AGN at $z=0.320$ (Bade et al. 1995).

{\it \ax{144109$+$3520(NO32)}} --- There is a $ROSAT$ HRI source
in the error circle of the AMSSn source. The optical counterpart
of the HRI source (34\_02) is identified with a narrow-emission line galaxy
at $z=0.077$. The line ratios of the narrow-emission lines are
AGN-like (see Figure~\ref{line_rat}). 36\_03 is an elliptical galaxy
at the same redshift. The redshift difference of the two galaxies
corresponds to the velocity difference of $1000$ km s$^{-1}$.

{\it \ax{144301$+$5208(NO26)}} --- There is a $ROSAT$ HRI source
at just outside of the error circle of the AMSSn source. 
The optical counterpart of the HRI source is identified with 
an Me-type star. The HRI source is faint, and if 
typical X-ray spectrum of stars (\apppi\ $> +2$) is assumed,
the expected hard band flux is well below the 
AMSSn survey limit. There is another $ROSAT$ HRI source 
$1.\!^{\prime}6$ south-west from the center of the AMSSn 
source. It is 3 times fainter than the former one.
A FIRST radio source in the error circle of the
AMSSn source is identified with a narrow-emission line galaxy. 
The large [OIII] to H$\beta$ flux ratio is consistent with 
narrow-line AGN. 
Considering that the X-ray spectrum of the AMSSn source is hard 
(\apppi\ $=+1.01\pm0.18$) and the HRI sources are faint,
The AMSSn source is identified with the narrow-line AGN.

{\it \ax{145026$+$1857(NO33)}} --- There is a $ROSAT$ PSPC source
just outside of the error circle of the AMSSn source. 
There is a clustering of galaxies around the PSPC source, 
and the X-ray spectrum of the
AMSSn source is soft (\apppi\ $=+2.17\pm0.17$). 
Thus the AMSSn source is identified with a cluster of galaxies.
The redshift of the cluster is not available.

{\it \ax{150339$+$1016(NO05)}} --- There is a $ROSAT$ PSPC
source in the error circle of the AMSSn source. In the PSPC
error circle, a FIRST radio source is detected. The optical
counterpart of the FIRST source is identified with a narrow-emission line
galaxy at $z=0.095$. The line ratios of the narrow-emission lines
are consistent with those of AGNs (Figure~\ref{line_rat}).
The radio source shows core and lobe structure.

{\it \ax{150423$+$1029(NO04)}} --- There is
a $ROSAT$ HRI source in the error circle of the AMSSn source.
The optical counterpart of the HRI source is a broad-line AGN 
at $z=1.839$ (Wilkes et al. 1983; PKS1502$+$106). 

{\it \ax{150430$+$4741(NO12)}} --- There is a $ROSAT$ HRI source
in the error circle of the AMSSn source. The optical counterpart
of the HRI source (13\_01) is identified with a broad-line AGN at $z=0.822$.
It is also detected in the NVSS and FIRST radio surveys.

{\it \ax{151441$+$3650(NE10)}} --- There is a $ROSAT$ HRI source
in the error circle of the AMSSn source. The optical counterpart
of the HRI source (13\_02) is
a broad-line AGN at $z=0.371$ (Schmidt 1974; 4C+37.43). 

{\it \ax{151524$+$3639(NO21)}} --- There are three bright optical objects
in the error circle of the AMSSn source. 12\_03 is a narrow-emission line
galaxy at $z=0.324$. The line ratios of the narrow-emission lines 
are AGN-like (see Figure~\ref{line_rat}). 
13\_01 is an elliptical galaxy at $z=0.3$. 
15\_02 is an M-type star. 13\_01 could be a companion galaxy
of 12\_03.

{\it \ax{155810$+$6401(NO22)}} --- There is a $ROSAT$ HRI source
in the error circle of the AMSSn source. The optical counterpart
of the HRI source (12\_01) is identified with a broad-line AGN at $z=0.352$.

{\it \ax{160118$+$0844(NO53)}} --- There is a $ROSAT$ HRI source
in the error circle of the AMSSn source. The optical counterpart
of the HRI source (AA) is a narrow-emission line object at $z=0.606$.
Considering large [OIII] to H$\beta$ flux ratio and hardness
of the X-ray spectrum (\apppi\ $=+0.93\pm0.32$), 
the object is identified as a narrow-line AGN.

{\it \ax{163538$+$3809(NO47)}} --- There is a $ROSAT$ PSPC
source in the error circle of the AMSSn source. The brightest object
in the PSPC error circle (10\_01) is identified with a 
narrow-emission line galaxy at $z=0.099$. The line ratios of the
narrow-emission lines are AGN-like (Figure~\ref{line_rat}). 

{\it \ax{163720$+$8207(NE08)}} --- There is a $ROSAT$ PSPC
source in the error circle of the AMSSn source. There are two
optical objects in the PSPC error circle. 22\_07 is a narrow-emission line
galaxy with HII-region like emission line ratios. AA shows no emission line.
A bright galaxy outside of the PSPC error circle 
(32\_08) is a broad-line AGN at $z=0.041$. 22\_07 is a companion galaxy
of 32\_08 with a velocity difference of 200 km s$^{-1}$.

{\it \ax{164045$+$8233(NE07)}} --- There is a cluster of galaxies
(Vikhlinin et al. 1998; No.183) at $z=0.26$. The X-ray spectrum of the
AMSSn source is soft (\apppi\ $=+2.05\pm0.18$)
and consistent with those of clusters of galaxies. Thus, 
the AMSSn source is identified with the cluster of galaxies.

{\it \ax{170305$+$4526(NO35)}} --- There is a $ROSAT$ HRI source
in the error circle of the AMSSn source. The optical counterpart of
the HRI source (34\_02) is identified with a broad-line AGN at $z=0.171$.

{\it \ax{170548$+$2412(NE11)}} ---  There is a $ROSAT$ HRI
source in the error circle of the AMSSn source. The optical
counterpart of the HRI source is a merging galaxy (20\_05 and 20\_04).
20\_05 is a broad-line AGN at $z=0.114$ (Stocke et al. 1991). 
20\_04 is a narrow-emission line galaxy with AGN-like 
line ratios (see Figure~\ref{line_rat}). 
Considering the \oiiia\ flux of 20\_05 is 3 times larger
than that of 20\_04 and the HRI source position is closer to 20\_05 than
20\_04, the AMSSn source is identified with
the broad-line AGN, 20\_05.

{\it \ax{170730$+$2353(NO24)}} --- This source is observed
with {\it Chandra}. The optical counterpart of the {\it Chandra}
source is identified with a narrow-emission line galaxy at $z=0.245$.
The line ratios of the narrow-emission lines are AGN-like 
(Figure~\ref{line_rat}).

{\it \ax{171125$+$7111(NO02)}} --- There are two $ROSAT$ HRI
sources around the error circle of the AMSSn source. 
The south-eastern HRI source is two times brighter than the
north-western one, and is detected also by RASS. Thus the 
AMSSn source is identified with the south-eastern HRI source.
The optical counterpart of the south-eastern source (41\_00) 
is a broad-line AGN at $z=1.011$ \citep{app98} and that of the 
north-western one (21\_08) is an M-type star. 

{\it \ax{171811$+$6727(NE21)}} --- There is a RASS source
in the error circle of the AMSSn source. An optical object
near the RASS center position (49\_03) is identified with 
a broad-line AGN at $z=0.549$.

{\it \ax{172815$+$5013(NO23)}} --- There is a $ROSAT$ HRI
source in the error circle of the AMSSn source. The 
optical counterpart of the HRI source (17\_01) is IZw187.
It is a BL Lac object at $z=0.054$.

{\it \ax{172938$+$5230(NO10)}} --- There are a RASS and $ROSAT$
PSPC sources in the error circle of the AMSSn source. 
A bright optical object near the center of the RASS source (23\_01) is
identified with a broad-line AGN at $z=0.278$.

{\it \ax{174652$+$6836(NO42)}} --- There is a $ROSAT$ PSPC
source in the error circle of the AMSSn source. The optical
counterpart of the PSPC source is an AGN
(Kriss \& Canizares 1982; VIIZw 742, 1E1747.3$+$6836).
This is a merging galaxy with two knots. The southern knot is
a broad-line AGN at $z=0.063$. The northern
knot shows no emission line and optical continuum 
similar to that of an elliptical galaxy.
The redshift difference of the two knots
corresponds to a velocity difference of
$630$ km s$^{-1}$.

{\it \ax{174943$+$6823(NO43)}} --- There is a $ROSAT$ PSPC
source in the error circle of the AMSSn source. The optical
counterpart of the PSPC source (26\_01) is 
a narrow-emission line AGN at $z=0.051$ (Iwasawa et al. 1997). 
The emission line ratios of the narrow
emission lines are AGN-like.

{\it \ax{210738$-$0512(NO17)}} --- There is a $ROSAT$ HRI
source in the error circle of the AMSSn source. The optical 
counterpart of the HRI source (AA) is identified with a broad-line
AGN at $z=0.841$.

{\it \ax{230719$-$1513(SE35)}} --- There is a $ROSAT$ PSPC
source just outside of the error circle of the AMSSn source.
There is a cluster of galaxies at $z=0.111$ (Abell et al. 1989; Abell 2533) in
the PSPC error circle. The X-ray spectrum of the AMSSn source
is consistent with those of clusters of galaxies (\apppi\
$=+2.02\pm0.10$).

{\it \ax{230738$-$1526(SE14)}} --- There is a $ROSAT$ PSPC
source just outside of the error circle of the AMSSn source. 
The brightest optical object in the PSPC error circle (19\_04) is
identified with a broad-line AGN at $z=0.199$.

{\it \ax{232639$+$2205(NO48)}} --- A bright galaxy (19\_01) in the
AMSSn error circle is identified with a broad-line AGN at $z=0.151$.

{\it \ax{233200$+$1945(NO18)}} --- This source is observed with
{\it Chandra} and {\it XMM-Newton}. 
The optical counterpart of the {\it Chandra} source (AA) 
is a narrow-emission line galaxy
at $z=1.416$. High ionization lines of [NeIV]$\lambda$2424 and 
[NeV]$\lambda$3426 are detected. Thus the object is classified as 
an AGN. There is a hint of broad MgII emission line.

{\it \ax{233253$+$1513(NO14)}} --- A bright galaxy in the
AMSSn error circle (18\_01) is identified with a broad-line 
AGN at $z=0.215$. 21\_02 shows no emission line.

{\it \ax{234725$+$0053(NE23)}} --- This source is observed with
{\it XMM-Newton}. Three sources are detected near the {\it ASCA}
source position. The brightest source is more than 
three times brighter than the other sources in 2--10~keV band
and consistent with the {\it ASCA} flux, thus 
we regard the brightest source as the counterpart of the {\it ASCA} source.
The X-ray spectrum of the source shows strong Fe emission line
at redshift of $0.213\pm0.006$ (Ueda et al. in preparation).
Assuming the redshift and the intrinsic photon index of 1.7,
the X-ray spectrum is fitted with $N_{\rm H}=2.5\pm0.5\times10^{22}$ cm$^{-2}$.
The estimated intrinsic luminosity is $L_{\rm 2-10keV} = 1.7\times10^{44}$
erg s$^{-1}$. The object is heavily obscured luminous AGN. 
No optical spectroscopy has been conducted for the optical 
counterpart of the {\it XMM-Newton} source (26\_00). Other AMSSn
AGNs with similar redshifts and $N_{\rm H}$ are identified with
AGNs without broad H$\beta$ emission line (e.g. Figure~\ref{lum_nh}), 
the object would have similar
optical spectrum. In the later discussions, we treat the object
as a heavily obscured no broad H$\beta$ AGN.

{\it \ax{235541$+$2508(NE28)}} --- There is a bright G5 star with
a V magnitude of 8.7 in the error circle of the AMSSn source. 
The soft spectrum of the AMSSn source (\apppi\ 
$=+3.28\pm0.08$) and small
hard X-ray to optical flux ratio are consistent with
those of G-type stars.

{\it \ax{235554$+$2836(NO20)}} --- 
There is a $ROSAT$ PSPC source in the error circle of the 
AMSSn source. The brightest object in the PSPC error circle is
a broad-line AGN at $z=0.729$ (Olsen 1970; 4C+28.59). 
There is a $ROSAT$ PSPC source just outside of the AMSSn error 
circle. The PSPC source is a galactic star. The X-ray spectrum of
the AMSSn source (\apppi\ $=+1.61\pm0.18$) 
is too hard to be a galactic star, 
thus the AMSSn source is identified
with the broad-line AGN.

\subsection{Summary of Emission Line Diagnostics of the Observed Objects}

In order to evaluate the strength and the width of detected 
emission lines, a spectral fitting for emission lines 
with the $\chi^{2}$ minimization method is conducted. 
For the fitting, {\bf specfit} command in {\bf spfitpkg} package 
in the IRAF is applied. All the emission lines are fitted with
Gaussian profile. For H$\alpha$ and H$\beta$ lines,
we put one broad and one narrow components independently. 
We assume that the narrow H$\alpha$ (narrow H$\beta$) line has
the same velocity width as \niib (\oiiia).
FWHMs of line widths are deconvolved by the spectral resolution 
mentioned in Section 3.2. The results are summarized in Table~\ref{id_line}.
In the table, equivalent widths of \oiiia, broad \hb,
narrow \hb, \niib, broad \ha, and narrow \ha, and
velocity widths of narrow \oiiia, broad \hb, and
broad \ha\ are summarized.
If the best-fit line flux is 3 times more than the estimated 
uncertainty of the line flux of a component, 
the best-fit line flux and the uncertainty is listed, 
otherwise one sigma upper limit for a component is listed. 
The upper limits for broad lines and narrow lines
are determined by assuming velocity width of 
3000 km s$^{-1}$ and 500 km s$^{-1}$, respectively.

Two thirds of the AMSSn sources (60/87) have 
optical counterparts
with a significant ($>3 \sigma$, except 
for \ax{134412$+$0016(NE20)}, see Section 3.3) broad emission 
line either in \ha, \hb, \mgii, 
\ciii, \civ, or \lya.
The velocity width of the broad H$\beta$ emission line
versus the \apppi\ is plotted in 
Figure~\ref{width_pi}. There is no population of 
narrow-line Seyfert 1s which show small H$\beta$ line 
width ($< 2000$ km s$^{-1}$) and large \apppi\
($> 2$) \citep{bra97}. 

Optical counterparts of 15 AMSSn sources only show 
narrow-emission lines whose line ratios are consistent
with those of Seyfert galaxies.
For 9 of them, both H$\alpha$ and
H$\beta$ wavelength ranges are observed.
The distribution of the \niib\ to H$\alpha$ and \oiiia\ to H$\beta$
ratios of the 9 objects is shown in 
Figure~\ref{line_rat} (solid
squares).\footnote{
In the diagram, we plot narrow-emission line galaxies 
which are spectroscopically observed, but not identified with
an X-ray source (cross). They are mostly serendipitous galaxies
with HII-region like narrow-emission line ratios. Two objects,
53\_07 of \ax{142651$+$2619(NO51)} and 20\_04 of \ax{170548$+$2412(NE11)}
have line ratios consistent with Seyfert galaxies. 
Since the two X-ray sources are identified with broad line AGNs,
we regard 53\_07 and 20\_04 to be serendipitous narrow line AGNs.}
The regions enclosed by the solid lines represent 
the regions occupied by
Seyfert galaxies, LINERs, and HII-region like galaxies
in the diagram \citep{vei87}.
All 9 objects fall in the region of Seyfert galaxies.
As for the 6 remaining objects,
\ax{174943$+$6823(NO43)} is identified with a narrow-line AGN
(Seyfert) by \citet{iwa97}, and detailed discussion on the narrow-line
ratios of the object, see the paper.
\ax{233200$+$1945(NO18)} shows high ionization narrow lines 
of [NeIV]$\lambda$2424 and [NeV]$\lambda$3426, thus we identify
the object with a narrow-line AGN. It also shows a hint of a
broad MgII$\lambda$2800 emission line. 
For the remaining 4 narrow-line galaxies, 
only \oiii\ narrow-emission lines are detected.
The logarithmic lower limits on \oiiia\ to H$\beta$ flux ratio are 
1.0 for \ax{123605$+$2613(NE04)},
1.1 for \ax{144301$+$5208(NO26)},
1.4 for \ax{151524$+$3639(NO21)}, and \\
1.5 for \ax{160118$+$0844(NO53)}.
The lower limits are consistent with Seyfert-like line ratios.
All of the identified narrow-line galaxies
are Seyfert-like galaxies, and there is no candidate of an
optical counterpart with LINER-like emission line ratios. 

Narrow-emission line ratios of broad-line 
AGNs with strong narrow-emission lines are also plotted with 
open squares. 
Most of the broad-line AGNs also have narrow-line
ratios similar to Seyfert galaxies, as we expect. \ax{111518$+$4042(NO56)} \\
and \ax{140532$+$5055(NO13)} have small \oiiia\ to H$\beta$
flux ratios and their line ratios fall amongst those of LINERs.
Since they have the narrower broad H$\beta$ lines than other
broad-line AGNs in the diagram, contamination by
broad H$\beta$ line possibly causes the small \oiiia\ to H$\beta$
line ratios.

\subsection{Reliability of the Identifications}

All but one of the 87 AMSSn hard X-ray selected sources are 
optically identified 
with either AGNs (including broad-line AGNs, narrow-line AGNs, 
and BL Lac objects (3)), clusters of galaxies (7), or 
a galactic star (1). Distribution of the distances between 
the optical counterparts and the X-ray sources 
normalized by 90\% error radius is shown in Figure~\ref{coord_orig_hist}.  
90\% of the optical counterparts are located within the 
estimated 90\% error radius, thus the distribution of the optical 
counterparts is consistent with the estimated uncertainties 
of the AMSSn X-ray source positions.

Based on optically-selected broad-line AGN counts \citep{har90}, 
the number of broad-line AGNs which 
fall in the error circles by chance can be estimated.
The total uncertainty area that is sum of the
areas of the error circles of the AMSSn sources
except for the sources whose error circles are tightly constrained
by either {\it Chandra} or {\it ROSAT} HRI is 
167 arcmin$^{2}$.
Most of the identified broad-line AGNs
are brighter than $R$ of 19 mag corresponding to 
$B$ of 19.5 mag for a typical QSO color ($B-R=0.5$ mag).
The cumulative number density of broad-line AGNs brighter 
than $B$ of 19.5 mag with redshift smaller than 2.2 is 
8.5 degree$^{-2}$ \citep{har90}. Thus, the expected number 
of contamination by broad-line AGNs is 0.4. 
It should be noted that positional information
from {\it ROSAT} PSPC and NVSS and FIRST radio sources
is used for a part of the sources, thus the number 
should be smaller. Therefore we conclude that
the contamination by chance coincidence is negligible.

For 15 AGNs showing only narrow-emission lines, 
11 of them are observed by either {\it Chandra}, {\it ROSAT} HRI,
NVSS, or FIRST, and the identifications are reliable. 
The remaining 4 objects are brighter than 18.5 mag
in $R$ band, and they show strong \oiii\ emission lines.
The number density of field galaxies at the magnitude is about 
100 degree$^{-2}$ (see e.g, Yasuda et al. 2001). 
About 40\% of the field galaxy population show strong \oiii\ 
emission lines at $B$ of 19 mag \citep{mad02} 
(it corresponds roughly $R$ of 18 mag), and 17\% of field 
emission line galaxies show AGN-like emission lines \citep{tre96}.
Therefore, the number density of AGN-like narrow-emission line 
galaxies brighter than $R$ of 18.5 mag is estimated to be 7 degree$^{-2}$, and 
the expected number of contamination of such galaxies in 167 arcmin$^{2}$
field is 0.32, and it is negligible.

There are 7 candidates of a cluster of galaxies in the 
AMSSn sample. Redshifts of the 6 candidates are
known, and they range from 0.073 to 0.260.
X-ray spectra of all of the 7 clusters of galaxies
are consistent with a soft X-ray spectrum of clusters of galaxies.
With the number of clusters of galaxies ($7\pm2.6$) and
the survey area (Section 2), 
the number density of clusters of galaxies above 
$3\times10^{-13}$ \ergs\ in the 2--10~keV band is 
estimated to be $\log N(>S) = -0.9\pm0.2$ degree$^{-2}$.
It is consistent with that determined in the ALSS 
with 2 clusters of galaxies at the same flux limit
($\log N(>S) = -0.9^{+0.2}_{-0.5}$ degree$^{-2}$; AOY00). 
This consistency supports the idea that 
the identifications of clusters of galaxies are also reliable.

\subsection{Comparison with Other Hard X-ray Selected AGN Samples}
The hard X-ray to optical flux ratio is one of
basic properties which reflect the nature of the 
detected objects. In Figure~\ref{opt_x1},
the optical magnitudes and the 2--10~keV hard X-ray fluxes
of the AMSSn AGNs (open circle) are compared with those of 
hard X-ray selected AGN samples from {\it HEAO1} A2 survey 
(open triangle; Piccinotti et al. 1982), ALSS (open square; AOY00), 
{\it Beppo-SAX} HELLAS survey (small dots; La Franca et al. 2002)
and {\it Chandra} survey of CDFN (Brandt et al. 2001; Barger et al. 2002)
and of CDFS \citep{gia02} (asterisks and filled triangles for upper limits).
The HELLAS survey is done in the 5--10~keV
band, and the flux in the energy band is converted
to the 2--10~keV flux, with
the photon index of typical AGNs ($\Gamma = 1.7$; e.g. Inoue 1985).
The 2--8~keV band is used as the hard band in the CDFN survey,
the 2--8~keV flux is converted to the 2--10~keV flux with
the effective photon index of each source. 
The {\it Chandra} objects do not necessarily show
AGN-like emission lines in the optical wavelength, 
but their hard X-ray emissions are thought to originate from AGNs.
The X-ray to optical flux ratio is determined by
$\log f_{\rm 2-10~keV}/f_{\rm R} = \log f_{\rm 2-10~keV} + R/2.5 +5.5$
which is derived with Kron-Cousin $R$ band response function \citep{hor01}.
The distribution of hard X-ray to optical flux ratios 
of AMSSn AGNs is similar to those of other hard X-ray 
selected samples.
AMSSn, ALSS and HELLAS samples have similar flux limits and
their ranges of X-ray to optical flux ratio are consistent with
each other ($\log f_{\rm 2-10~keV}/f_{R} = -2 - +2$). 
The range is 
larger than that of {\it HEAO1} A2 sample ($\log f_{\rm 2-10~keV}/f_{R} = -0.5 - +1$).
The deep {\it Chandra} samples have a larger range than 
the {\it ASCA} samples ($\log f_{\rm 2-10~keV}/f_{R} = -4 - +3$).
A significant number of AMSSn AGNs have hard X-ray
to optical flux ratios as large as the optically-faint hard X-ray
source seen in deep {\it Chandra} 
surveys ($\log f_{\rm 2-10~keV}/f_{R} > +1$). 
The optically-faint objects are candidates of high-redshift
absorbed AGNs (Section 4.2.2).

As mentioned in Section 1,
the distribution of the AMSSn AGNs in the
redshift versus hard X-ray luminosity diagram
is different from deep {\it Chandra} sample of AGNs.
In order to compare the distributions of hard
X-ray selected AGNs, the redshift and hard X-ray 
luminosity of the AMSSn, ALSS (AOY00), 
{\it HEAO1} A2 \citep{pic82}, and {\it Chandra} 
CDFN (Brandt et al. 2001; Barger et al. 2002) samples of AGNs
are plotted in Figure~\ref{amss_lum4}. From the CDFN sample, only 
spectroscopically identified objects are plotted. 
The 2--10~keV intrinsic luminosity,
$L_{\rm 2-10keV}$, of each AMSSn AGN is calculated by 
correcting for the redshift ($k$-correction), 
the Galactic absorption, and the 
estimated intrinsic absorption to the nucleus.
The intrinsic absorption is estimated 
from the hardness ratio corrected for 
Galactic absorption and the instrument response,
assuming that intrinsic nuclear emission of AGNs 
is a power-law spectrum with photon index of 1.7 (Section 4.1). 
For AGN with apparent X-ray spectra softer than 
\apppi\ of 1.7, the luminosity is calculated with
the best fit power-law index. The estimated luminosities are 
listed in Table~\ref{id_lum}.
The 2--10~keV luminosities of the CDFN objects are 
calculated in the same way as those for AMSSn AGNs 
using \apppi\ in the 0.5--8~keV, count rate in the 
hard band, and response function of {\it Chandra} ACIS.
The intrinsic photon index is assumed to be 1.7 (Section 4.1).
The redshift range of the 
AMSSn and ALSS samples of AGNs is similar to that of CDFN 
sample, but their luminosity is typically 2 orders of magnitude 
larger than CDFN sample at the same redshift.
The AMSSn sample consists of Seyfert galaxies 
($L_{\rm 2-10keV} < 1\times10^{44}$ \erg) at redshifts around 0.1 and 
QSOs ($L_{\rm 2-10keV} > 1\times10^{44}$ \erg) at higher redshifts up to 2. 
Thus {\it ASCA} selected samples of AMSSn and ALSS AGNs 
are suitable to study
luminous AGNs (QSOs)
in the intermediate redshift universe.

\section{ABSORPTION TO THE NUCLEUS IN THE HARD X-RAY SELECTED AGNS}

\subsection{Absorption to the Nucleus Estimated by X-ray Hardness}

A significant fraction of the identified AGNs 
in the AMSSn and ALSS show hard X-ray spectra.
The redshift versus \apppi\ diagram of the identified
AGNs is shown in Figure~\ref{z_pi} along with ALSS AGNs. 
More than 80\% of the AMSSn and ALSS AGNs lie \apppi\ between 1 and 2.5, 
but 11 objects have \apppi\ smaller than 1 in the 
AMSSn and ALSS AGNs. 
Absorption to their nuclei is thought 
to make their X-ray spectra very hard.
With the X-ray spectrum of each source, the
amount of the X-ray photoelectric absorption  
to the source can be estimated by assuming the
intrinsic spectrum and adopting the redshift of
the source. 

Distributions of 0.7--10~keV apparent photon index, \apppi\, 
of the AMSSn and ALSS AGNs 
are shown in Figure~\ref{pi_hist}. 
The \apppi\ distributions of AMSSn and ALSS samples 
above \apppi\ of 1 are consistent with a Gaussian 
distribution with an average photon index of 1.7 
and a sigma of 0.2 comparable to the uncertainties of 
the \apppi\ determinations for objects with
\apppi\ $>1$ (Table~\ref{id_flux}). The average photon index is
consistent with that determined in the CDFN survey ($\Gamma=1.7$ in the
0.5--8~keV band; Barger et al. 2002),
and in the {\it XMM-Newton}
observation of Lockman Hole AGNs ($\Gamma=1.9\pm0.9$ in the
0.5--10~keV band; Mainieri et al. 2002).
Therefore a power-law with
photon index of 1.7 is used as the intrinsic X-ray spectrum of AGNs.

The conversion from the \apppi\ to the
amount of the absorption depends on the redshift,
because higher-energy photons of the source frame is 
observed at higher redshift.
In Figure~\ref{z_pi}, the upper and lower solid lines 
in the figure correspond to the \apppi\ of an object at 
the redshift with the intrinsic photon index of 1.7 and 
X-ray absorption with hydrogen column densities, $N_{\rm H}$, 
of $1\times10^{22}$ cm$^{-2}$ and  $1\times10^{23}$ cm$^{-2}$, respectively.
The X-ray sources with \apppi\ smaller than 1 
correspond to AGNs at intermediate redshifts affected by 
X-ray absorption with $N_{\rm H}$ of $1\times10^{22}$ 
cm$^{-2}$ $-$ $1\times10^{23}$ 
cm$^{-2}$, and high-redshift AGNs with 
$N_{\rm H} > 1\times10^{23}$ cm$^{-2}$.

The estimated absorption is listed in Table~\ref{id_lum},
and plotted as a function of absorption-corrected luminosity in 
Figure~\ref{lum_nh} 
for $z<0.6$(a) and $z>0.6$(b) AGNs, along with ALSS AGNs respectively. 
At $z<0.6$, five luminous ($L_{\rm 2-10keV} > 1\times10^{44}$ 
erg s$^{-1}$) absorbed ($N_{\rm H} > 1\times10^{22}$ cm$^{-2}$) AGNs 
are found in the AMSSn sample. 
At $z>0.6$, there are two AGNs with $L_{\rm 2-10keV} >  
1\times10^{45}$ erg s$^{-1}$ 
and $N_{\rm H} > 1\times10^{23}$ cm$^{-2}$. 
These objects are candidates
of long-sought absorbed luminous AGNs (type-2 QSOs). 

\subsection{Relation between X-ray and Optical Properties}

The large column density of absorbing material
($>1\times10^{22}$ cm$^{-2}$)
inferred from the hard X-ray spectrum can
contain large amount of dust, and should
affect the optical properties of the AGNs, 
i.e. apparent strength of broad lines,
faintness of the optical continuum, and
redness of the optical and near-infrared continuum.
The strength of broad lines, the X-ray-to-optical 
flux ratio, and the near-infrared
colors are discussed in the following subsections.

\subsubsection{Strength of Broad Lines}

If broad emission lines are present in all AGNs,
the non-detection of these lines should 
imply the presence of significant amount of dust
along our line-of-sight \citep{ant93}.
For more than half of the AMSSn and ALSS AGNs, 
that is all of the AMSSn and ALSS AGNs at $z<0.6$, 
H$\beta$ emission lines are measured
by our observation, thus
we use the emission line as an indicator.
In Figure~\ref{z_pi} and ~\ref{lum_nh}(a), 
AMSSn and ALSS AGNs without significant 
broad H$\beta$ line (no broad H$\beta$ AGNs) 
are marked with large crosses. 
Most of the AGNs with hard X-ray spectra do not
show broad H$\beta$ emission, and
the no broad H$\beta$ AGNs have 
significantly larger $N_{\rm H}$ than that of the 
broad H$\beta$ AGNs. The Kolmogorov-Smirnov test (K-S test)
indicates that the $N_{\rm H}$ distribution of the 
no broad H$\beta$ AGNs is different from that
of the broad H$\beta$ AGNs with 99.9\% confidence.

Among AGNs at $z>0.6$, 
\ax{160118$+$0844(NO53)} and \\ \ax{233200$+$1945(NO18)}
do not show strong broad emission lines, 
and strong narrow emission lines are detected.
$N_{\rm H}$ estimated from X-ray spectra are
$(3\pm2)\times10^{22}$ cm$^{-2}$ and $(10\pm5)\times10^{22}$ 
cm$^{-2}$. Therefore, they are candidates to be the 
luminous cousins of type-2 Seyferts at high-redshifts, 
i.e. type-2 QSOs.
The other high-redshift AGNs show strong broad emission 
lines of either \mgii, \ciii\, or \lya.
\ax{133937$+$2730(NE17)}, \ax{150423$+$1029(NO04)},
\\ and \ax{210738$-$0512(NO17)}
show strong broad emission lines,
although they are estimated to have
$N_{\rm H} >1\times10^{22}$ cm$^{-2}$.
The $N_{\rm H}$ corresponds to an extinction of more 
than 10 magnitude at the wavelength of \mgii\ with the
Galactic $N_{\rm H}$ to $A_{V}$ ratio
($N_{\rm H}/A_{V} = 1.79\times10^{21} {\rm cm}^{-2}$;
Predehl \& Schmitt 1995) and the Galactic 
extinction curve ($A_{\rm 2800\AA}/A_{V} = 1.9$ ;
Cardelli, Clayton, \& Mathis 1989). 
The broad emission lines would not be detected
through such a large absorbing column, 
implying a different line of sight to the X-ray
source, dust-free material (e.g. dust sublimation in
warm absorbing gas) or different dust properties 
(e.g. different dust size distribution in AGNs;
Maiolino et al. 2001a,b).
Similar AGNs with broad emission lines and
large $N_{\rm H}$ are summarized in Maiolino et al. (2001a).
There is another possibility that their optical nuclei are 
obscured and the broad lines come from scattered nuclear light
(Aligheri et al. 1994; Akiyama \& Ohta 2001).
It is also possible that their intrinsic X-ray spectra
are harder than a power-law with a photon index of 1.7
and the estimated large amount of X-ray absorption is not real.
A reflection component is often present in Seyfert galaxies above 6~keV
and may explain the hardening of the X-ray spectrum 
of high-redshift QSOs (AOY00).

\subsubsection{Hard X-ray to Optical Flux Ratio}

AGNs which have large hard X-ray to 
optical flux ratio ($\log f_{\rm 2-10~keV}/f_{R} > +1$)
are marked with small dots in Figure~\ref{z_pi} and
~\ref{lum_nh}. Six out of 10 of the optically-faint 
AMSSn and ALSS AGNs are at $z>0.6$. In the redshift range,  
the objects with \apppi\ $<$ 1
are optically faint and X-ray bright and
are consistent with absorption in the optical/UV.
Three of the broad-line AGNs with 
$N_{\rm H} > 1\times10^{22}$ cm$^{-2}$ mentioned
in the previous subsection also have large
hard X-ray to optical flux ratio suggesting
dust absorption in the optical continuum emission.  
It should be noted that 5/10 of the optically-faint
AGNs are radio-loud AGNs. It is consistent with
the fact that radio-loud AGNs have larger X-ray to
optical flux ratio on average \citep{zam81}.

\subsubsection{Near-infrared Color}

The near-infrared color of an AGN can be another indicator
of the amount of dust absorption to the nucleus.
About two thirds of the AMSSn AGNs (48/76) and all of the 
ALSS AGNs are covered in the area of the second incremental 
database of the Two Micron All Sky Survey
(2MASS; Kleinmann et al. 1994). The 2MASS is an all sky survey
in the $J$-, $H$-, and $K_{S}$- bands down to $J$ of 16.5 mag, 
$H$ of 15.5 mag, and $K_{S}$ of 15.0 mag.
In the covered AGNs, 
$56\pm6$\% (44/78) of them are detected in the survey. The $J$-,
$H$-, and $K_S$-bands magnitudes of detected AMSSn AGNs
are listed in Table~\ref{id_flux}.
The $J-K_S$ colors of the AMSSn and ALSS
AGNs are plotted as a function of redshift
in Figure~\ref{z_jk}. The $J-K_{S}$ colors of optically-selected 
AGNs \citep{bark01} and {\it HEAO1} A2 AGNs \citep{kot92}
are also plotted in the same figure with small dots and small crosses,
respectively. 
At redshifts smaller than 0.3, elliptical galaxies (solid line,
no evolution is considered) are
bluer than AGNs, thus the contribution from the host
galaxy of an AGN make its color bluer than a typical AGN.
At redshift smaller than 0.2, the color distribution of
the AMSSn and ALSS AGNs is bluer and closer to the 
track of an elliptical galaxy than that of {\it HEAO1} A2 AGNs on average.
Above this redshift, the color range of the AMSSn AGNs is
similar to that of optically-selected AGNs. 

The criteria of $J-K_{S}$ color
redder than 2 is used for selecting red QSOs in the 2MASS survey. 
There are four AGNs with $J-K_S$ color 
redder than 2 in the AMSSn AGNs. No AGNs in the ALSS sample
have $J-K_S$ color redder than 2.
The $J-K_{S}$ colors ($\le 3$) and redshift range ($z\le0.7$) of 
the red AGNs are similar to those of the 2MASS red QSO 
population \citep{wil02}. 
The $J-K_S$ color of \ax{121854$+$2957(NO07)} is very red: 
1.0 magnitude redder than the average of optically-selected AGNs, 
and 1.8 magnitude redder than the bluest AMSSn AGN at the redshift
of 0.2. If it is assumed that the dust absorption to the nucleus causes
the red $J-K_S$ color of the object, and intrinsic $J-K_S$ color
is similar to those of 2MASS and AMSSn AGNs at the same redshift
($1.2 < J-K_S < 2.4$), the amount of dust absorption 
to the nucleus is estimated to be $A_V$ of $3-9$ mag at the object frame with
the Galactic extinction curve. 
With the Galactic $A_V$ to $N_{\rm H}$ ratio,
the estimated amount of dust absorption can be converted to 
X-ray absorption column density of 
$(1-2)\times10^{22}$ cm$^{-2}$, which is roughly consistent with 
that calculated from the X-ray hardness [$(3\pm1)\times10^{22}$ cm$^{-2}$].
The amount of the X-ray absorption to the nucleus is also
similar to those observed in 2MASS red AGNs \citep{wil02}.

At $z>0.3$, large fraction of AMSSn and ALSS AGNs 
are not detected in the 2MASS survey (plotted at $J-K_S<0$ in 
Figure~\ref{z_jk}), because the 2MASS 
survey limit is not deep enough to detect them. We conducted 
near-infrared photometric observations of
AMSSn and ALSS AGNs not detected in the 2MASS survey
to assess the fraction of red QSOs. 
The results will appear elsewhere.

\subsection{Luminosity Distributions of Absorbed and Unabsorbed AGNs}

The luminosity distribution of
less-absorbed ($N_{\rm H}<1\times10^{22}$ cm$^{-2}$) 
and absorbed ($N_{\rm H}>1\times10^{22}$ cm$^{-2}$) AGNs
are compared in Figure~\ref{Ldis_B}.
We combined samples of AMSSn, ALSS and {\it ASCA} Deep 
Survey in the Lockman Hole (ADSL; Ishisaki et al. 2001).
ADSL sample consists of 12 hard X-ray selected AGNs
above flux limit of $3\times10^{-14}$ \ergs (Ishisaki et al. 2001).
Because at high redshifts, AGNs with $N_{\rm H} = 1\times10^{22}$
cm$^{-2}$ can not be distinguished from AGNs without absorption
(Figure~\ref{z_pi}), we limit the sample of AGNs up to 
redshift of 0.6. Below this redshift, H$\beta$ wavelength
is covered for all AGNs.
The fraction of the absorbed AGNs in the luminous AGNs with
$L_{\rm 2-10keV} > 3\times10^{45}$ \erg\ (1/13, $8\pm8$\%)
is smaller than that in the less-luminous AGNs with
$L_{\rm 2-10keV} < 3\times10^{45}$ \erg\ (14/70, $20\pm5$\%).
But, the K-S test on the luminosity distributions
of absorbed and less-absorbed AGNs indicates that 
there is no significant difference between the two 
luminosity distributions (86\% significance).

In order to extend the comparison to lower
intrinsic luminosity range, we combined 
the hard X-ray selected {\it Chandra} CDFN AGNs
(Barger et al. 2002). 
Spectroscopic redshifts are available only for 
objects brighter than $R$ of 24 mag, thus optical 
spectroscopic identification of CDFN is not complete, 
especially for high-redshift absorbed AGNs. The spectroscopic 
identification is also limited by the fact that strong 
emission lines go out of optical wavelength at redshifts 
larger than 1.4. For objects without 
spectroscopic redshifts, their photometric redshifts are used,
and we also limit the sample up to redshift of 0.6.
Intrinsic absorption and luminosity of the AGN is estimated
with the \apppi\ in the 0.5--8~keV by assuming an 
intrinsic photon index of 1.7 in the same way as those for AMSSn AGN.
The luminosity distributions of
less-absorbed and absorbed AGNs are compared 
combining the {\it ASCA} and CDFN samples. 
The less-absorbed AGNs marginally have larger
luminosity than absorbed AGNs; the K-S test shows
that the luminosity distributions of less-absorbed and
absorbed AGNs are marginally different with 98\% confidence.
The marginal differences of luminosity
distribution of absorbed and less-absorbed AGNs may
indicate that the fraction of absorbed AGNs decreases
with increasing intrinsic luminosity.

In order to evaluate the {\it intrinsic} fraction
of absorbed AGNs, the detection limit of 2--10~keV
selection has to be corrected. In the AMSSn AGNs, 
since the sample is limited by the count rate, the limit to 
the intrinsic luminosity at each 
redshift depends on the amount of absorption. 
Detailed evaluation of the luminosity 
function of less-absorbed and absorbed AGNs 
and the estimation of the 
{\it intrinsic} fraction of absorbed AGNs
are discussed in a separate paper (Ueda et al. 2003). 

As summarized in Section 4.2.1, the non-detection
of the broad H$\beta$ emission line can be used
as an indicator of dust absorption to the nucleus.
The luminosity distribution of
broad H$\beta$ and no broad H$\beta$ AGNs
in the AMSSn, ALSS, and ADSL are compared in Figure~\ref{Ldis_C}.
The luminosity distribution of the broad H$\beta$ AGNs
is also marginally different from that of the no broad H$\beta$ AGNs.
K-S test indicates that 
the broad H$\beta$ AGNs have larger intrinsic luminosity
than the no broad H$\beta$ AGNs with 97\% confidence.
It is suggested that the fraction of no broad H$\beta$ 
AGNs in QSOs is smaller than that in the
Seyfert galaxies. Such luminosity dependence
of the fraction of narrow-line AGNs is also
reported in a radio selected sample of AGN
(e.g. Lawrence 1991) and a far-infrared
selected sample (e.g. Barcons et al. 1995).

\section{RADIO PROPERTIES OF THE AMSSN AGNS}

Radio loudness is one of important properties of AGNs.
$37\pm5$\%(29/79) AMSSn AGNs including 3 BL Lac objects
are detected either in the NVSS or
FIRST survey. The detection rate is similar to that in the 
ALSS ($33\pm9$\%; 10/30). The optical and radio 
luminosities, and hard X-ray and 
radio luminosities of the AMSSn AGNs (circles) are plotted in 
Figures~\ref{radio_opt1} and ~\ref{radio_x1}.
The 1.4GHz radio luminosity of each source is calculated
from the 1.4 GHz NVSS flux of the source
assuming a power-law radio spectrum with index of $-0.5$.
If the FIRST flux is available, the flux is used for the calculation.
For AGNs without NVSS detection, the upper limits
on the radio luminosities are calculated with
the detection limit of NVSS (2.5mJy), and
indicated with downward arrows. ALSS AGNs (squares) are also
plotted in the figures. In Hooper et al. (1996), AGNs above the 
dashed line in Figure~\ref{radio_opt1} are defined as 
radio-loud AGNs ($R_{8.4} \equiv \log L_{8.4} / L_{B} > +1$).
$L_{8.4}$ and $L_{B}$ of AMSSn and ALSS AGNs are derived from 
$L_{1.4}$ and $M_{V}$ by assuming power-law index of $-0.5$.
Following the radio-loudness criterion, the number of radio-loud 
AGNs is 20 excluding AGNs with radio upper limit, and the fraction is
$18\pm4$\% (20/109) in the AMSSn and ALSS AGNs.

The optical luminosity is affected by dust absorption,
and the radio to optical luminosity ratio may not be a good indicator
of intrinsic radio-loudness. In Figure~\ref{radio_x1},
the AMSSn AGNs satisfying the above radio-loud criterion are marked 
with large squares.
Typical hard-X-ray to radio 
luminosity ratio in QSOs are shown with dashed lines for 
radio-loud QSOs (upper) and radio-quiet QSOs (lower) (Elvis et al. 1994).
In the distribution of the AMSSn and ALSS AGNs, 
there is a clear gap of 1 order of magnitude below the
radio-loud QSO line. There are four AGNs located
within the scatter of radio-quiet AGNs,
but are identified as radio-loud AGNs from their 
radio to optical flux ratios.
If the sample is divided at the solid 
line, there are 17 radio-loud AGNs in the AMSSn and ALSS AGNs, 
thus the fraction
of radio-loud AGNs is $16\pm4$\% in total. The fraction is higher than that
in the {\it ASCA} LSS sample of AGNs (10\%, AOY00). 

The fraction of the radio-loud objects increases as the X-ray
luminosity increases; $10\pm5$\% (4/41) and $43\pm11$\% (9/21)
for AGNs with $3\times10^{41} < L_{\rm 2-10keV} < 1\times10^{44}$ \erg 
and $1\times10^{45} < L_{\rm 2-10keV} < 3\times10^{46}$ \erg,
if we define radio-loudness with radio to X-ray luminosity ratio.
A K-S test comparing the $L_{\rm 2-10keV}$ distribution of radio-loud objects 
and that of radio-quiet objects shows that the two distribution 
is different with $>$99.95\% confidence.
A similar tendency is also detected 
in the Palomar-Green Bright Quasar Survey and 
the Einstein Extended Medium Sensitivity Survey \citep{hoo96}.
In the EMSS sample of AGNs, all but one radio-loud AGNs have
0.3--3.5~keV luminosities, $L_{\rm 0.3-3.5keV} > 1\times10^{44}$ erg s$^{-1}$, 
even though many radio-quiet QSOs have $L_{\rm 0.3-3.5keV} <
1\times10^{44}$ erg s$^{-1}$ \citep{del94}.
The tendencies are consistent with the fact that
radio-loud QSOs have 3 times larger hard X-ray luminosity
than radio-quiet QSOs at a given optical luminosity \citep{zam81}.

\section{SUMMARY}

We have presented the results of optical spectroscopic 
identifications of a bright subsample of hard 
X-ray selected sources from the AMSS in the
northern sky (AMSSn).
The total survey areas are 34 degree$^{2}$ and 68 degree$^{2}$
at flux limits of $3\times10^{-13}$ \ergs\ and 
$1\times10^{-12}$ \ergs\ in the 2--10~keV band, respectively.
All of the 87 hard X-ray selected sources are 
identified, with AGNs (including broad-line AGNs, 
narrow-line AGNs, and 3 BL Lac objects), 7 clusters 
of galaxies, and 1 galactic star. It is the largest 
complete sample of hard X-ray selected AGNs at the 
bright flux limit. The AMSSn AGNs consists of
Seyfert galaxies at redshifts around 0.1 and 
QSOs at higher redshifts up to 2. Their luminosities
are typically two orders of magnitude larger than 
the deep CDFN sample, and the AMSSn AGNs sample
is suitable to study luminous AGNs in the intermediate
redshift universe.

Amounts of absorption to the nuclei of the AGNs are
estimated at hydrogen column densities of up to
$\sim3\times10^{23}$ cm$^{-2}$ from their X-ray spectra. 
There are several
candidates of luminous ($L_{\rm 2-10keV} > 1\times10^{44}$ \erg)
absorbed ($N_{\rm H} > 1\times10^{22}$ cm$^{-2}$) AGNs,
i.e., type-2 QSOs.
Optical properties 
of X-ray absorbed AGNs with $N_{\rm H} > 1\times10^{22}$ cm$^{-2}$ 
indicate the effects of dust 
absorption. At $z<0.6$, most of 
the X-ray absorbed AGNs do not show broad H$\beta$ 
emission lines. At $z>0.6$, 
the X-ray absorbed AGNs show large hard 
X-ray to optical flux ratios ($\log f_{\rm 2-10~keV}/f_{R} > +1$).
However, three high-redshift AGNs with strong broad lines show
hard X-ray spectra with $N_{\rm H} > 1\times10^{22}$ cm$^{-2}$.
The column density corresponds to extinction of more 
than 10 mag at the wavelength of \mgii\ with 
Galactic $N_{\rm H}$ to $A_{V}$ ratio and Galactic 
extinction curve, and the strong broad lines would
not be detected. 

In combination with hard X-ray selected AGN samples
from the {\it ASCA} Large Sky Survey, the {\it ASCA}
Deep Survey in the Lockman Hole and CDFN, 
the luminosity distribution of absorbed ($N_{\rm H} > 1\times
10^{22}$ cm$^{-2}$) and less absorbed ($N_{\rm H} < 1\times
10^{22}$ cm$^{-2}$) AGNs are compared at $z<0.6$. 
There is a marginal difference of luminosity
distributions of absorbed and less-absorbed AGNs, it may
indicate that the fraction of absorbed AGNs decreases
with increasing intrinsic luminosity.

\section{NOTE ADDED IN PROOF}
We conducted an optical spectroscopic observation of the
optical counterpart of 1AXG~J234725+0053(NE23) with FOCAS 
on Subaru telescope. The object shows AGN-like 
emission lines at $z=0.233\pm0.001$, which is slightly 
larger than that determined from the X-ray spectrum. 
No broad H$\beta$ emission line is detected.

A follow-up observation of 1AXG~J133937+2730(NE17) with 
XMM-Newton detected two X-ray sources in the error
circle. The coordinates are 
13$^{\rm h}$ 39$^{\rm m}$ 36$^{\rm s}$
$+$27$^{\circ}$ 30$'$ 48$"$ and
13$^{\rm h}$ 39$^{\rm m}$ 39$^{\rm s}$
$+$27$^{\circ}$ 30$'$ 24$"$. In the text, we identified
the ASCA source with the latter one. However,
the former one is 2.5 times brighter than the latter one
in the 2--10~keV band. The former
one is associated with a galaxy.

\acknowledgments

We would like to thank members of UH88$^{\prime\prime}$ telescope,
KPNO, and Subaru Telescope team, especially, Dr. Youichi Ohyama, 
who gave us opportunity to observe an object and 
advice on FOCAS observing run. We appreciate the help of
Drs. Akihiko Tomita and Ingo Lehmann for the spectroscopic
identification. We are grateful to anonymous referee for
invaluable comments. We thank Dr. Chris Simpson for helpful comments
that improved the manuscript.
This publication makes use of data products from the 
Two Micron All Sky Survey, which is a joint project of the 
University of Massachusetts and the Infrared Processing and 
Analysis Center/California Institute of Technology, funded 
by the National Aeronautics and Space Administration and 
the National Science Foundation.
This research has made use of the NASA/IPAC Extragalactic 
Database (NED) which is operated by the Jet Propulsion Laboratory,
California Institute of Technology, under contract with the 
National Aeronautics and Space Administration, and data obtained 
from the High Energy Astrophysics Science Archive Research 
Center (HEASARC), provided by NASA's Goddard Space Flight Center.

\clearpage

\begin{table}
\caption{
CANDIDATES OF OPTICAL COUNTERPARTS AND RESULTS OF THE
IDENTIFICATIONS
\label{id_table}
}
\end{table}

\begin{table}
\caption{
APPARENT PROPERTIES OF IDENTIFIED OBJECT
\label{id_flux}
}
\end{table}

\begin{table}
\caption{
ABSOLUTE PROPERTIES OF IDENTIFIED OBJECTS
\label{id_lum}
}
\end{table}

\begin{table}
\caption{
OPTICAL EMISSION LINE PROPERTIES OF IDENTIFIED OBJECTS
\label{id_line}
}
\end{table}

\clearpage

\figcaption{
Left) $3^{\prime}\times3^{\prime}$ finding
charts centered
on the positions of AMSSn sources. North is up and
east is to the left. A large circle at
the center of each finding chart represents 
the 90\% error region of AMSSn source.
(Only for 1AXG~J000605$+$2031(NE01), the center
of the finding chart is shifted.) 
Circles with radii of $10^{\prime\prime}$,
$30^{\prime\prime}$, and $45^{\prime\prime}$
represent positions of {\it ROSAT} HRI, PSPC, and
RASS sources, respectively. The triangles indicate
positions of {\it Chandra} and {\it XMM-Newton} sources. 
The small and large 
squares indicate positions of FIRST and NVSS
1.4GHz radio sources, respectively.
Middle)
Close up view of the optical counterpart
of each X-ray source is shown. The field of view is
$15^{\prime\prime}\times15^{\prime\prime}$. 
Right) Optical spectrum of the identified object.
The optical spectra are flux calibrated by standard star
observation.
The wavelengths of major emission and absorption lines
are marked with vertical lines. The hatched area represents
the wavelength affected by strong night sky lines or 
atmospheric absorptions.
\label{Plates}}

\figcaption{
Stacked surface number density distribution of 
NVSS 1.4 GHz radio sources centered on the 87 
AMSSn hard X-ray selected source positions.
The uncertainty is estimated by assuming
Poisson statistics in each bin.
There is an excess in the number
density of radio sources within $\sim 1^{\prime}$ of
AMSSn hard X-ray selected sources. 
\label{NVSS_dis}}

\figcaption{
H$\beta$ broad-line width versus 
apparent photon index (\apppi) for
AMSSn (open circles) and ALSS AGNs
(open squares).
There is no population of narrow-line Seyfert 1s which
show a small H$\beta$ line width ($\sim$ 1000 km$^{-1}$) 
and a large apparent photon index ($\Gamma_{\rm app} > 2$).
\label{width_pi}}

\figcaption{
\niib\ to H$\alpha$ narrow emission line ratio versus \oiiia\ to H$\beta$
narrow emission line ratio diagram for narrow lines in the AMSSn objects. 
Identified narrow-line AGNs are indicated with filled squares.
Narrow-line objects which are not identified with the X-ray sources,
i.e. serendipitous objects, are marked with crosses. The 
open squares represent broad-line AGNs with strong narrow-lines in 
the AMSSn sample. The region occupied by Seyfert galaxies, LINERs, and 
HII-region like galaxies are shown with solid lines 
(Veilleux and Osterbrock 1987).
All of the identified narrow-line AGNs lie in the
region occupied by Seyfert galaxies.
\label{line_rat}}

\figcaption{
Distribution of the ratio between the 
distance of optical counterpart to the X-ray source
center and the estimated 90\% error radius of the 
X-ray source. The distribution is consistent with
the estimated 90\% error radius.
\label{coord_orig_hist}}

\figcaption{
$R-$band magnitude versus 2--10~keV hard X-ray
flux distribution of hard X-ray selected objects.
The AMSSn AGNs are plotted with open circles.
Open squares and small dots indicate ALSS AGNs (AOY00) and 
{\it Beppo-SAX} objects (La Franca et al. 2002),
respectively. Open triangles are {\it HEAO1} A2 AGNs (Piccinotti et al. 1982). 
Asterisks indicate sample from {\it Chandra} surveys in 
CDFN (Brandt et al. 2001, Barger et al. 2002) and CDFS 
(Giacconi et al. 2002). Upper limits
in the sample is indicated with filled triangles.  
Dashed lines represent the constant X-ray to optical flux
ratios of $\log f_{\rm 2-10keV} / f_R =$ $+3$, $+2$, $+1$, $0$, $-1$, $-2$, $-3$, 
and $-4$ from top to bottom.
The 2--10~keV hard X-ray fluxes of AMSSn AGNs are
determined from the count rates, the \apppi\ 
of each source, and GIS instrument response. 
Galactic absorption is corrected.
\label{opt_x1}}

\figcaption{
Hard X-ray luminosity versus redshift for
hard X-ray selected AGNs.
Open circles and open squares are AMSSn and
ALSS AGNs, respectively.
Open triangles are {\it HEAO1} A2 samples. 
Asterisks represent sample from {\it Chandra} survey in 
CDFN (Brandt et al. 2001, Barger et al. 2002). 
The {\it ASCA} AGN sample is two orders of magnitude
brighter than the {\it Chandra} objects at the same
redshifts.
\label{amss_lum4}}

\figcaption{
Apparent photon index in the 0.7--10~keV band
versus redshift for AMSSn (open circles) and
ALSS (open squares) AGNs.
The uncertainties of the \apppi\ determinations
are typically 0.2 (0.5) for objects with
\apppi\ $>1$ ($<1$).  
No broad H$\beta$ AGNs
are marked with large crosses.  
Small dots indicate objects with
$\log f_{\rm 2-10~keV}/f_{R} > +1$.
BL Lac objects are marked with large circles.
The horizontal dashed line indicates the \apppi\
of 1. The vertical dashed line marks the redshift of 0.6. 
The wavelength range of the H$\beta$ line is
observed only for objects below this redshift.
The solid lines show the \apppi\ of the intrinsic 
power-law continuum with $\Gamma$ of 1.7 absorbed 
by $N_{\rm H} = 
1\times10^{22}$ cm$^{-2}$ (upper) and $1\times10^{23}$ cm$^{-2}$ 
(lower) at each redshift. 
\label{z_pi}}

\figcaption{
Distribution of apparent photon index (\apppi)
of the AMSSn (shaded) and ALSS AGNs (open).
Both of the distributions peak at \apppi\ of 1.5 to 2.
\label{pi_hist}}

\figcaption{
Estimated hydrogen column density versus
absorption-corrected 
hard X-ray luminosity for the AMSSn (open circles) 
and ALSS AGNs (open squares) at $z<0.6$
(a) and at $z>0.6$ (b).
AGNs with \apppi\ $>1.7$ are 
plotted at $N_{\rm H} < 1\times10^{20}$ cm$^{-2}$
with uncertainty of 0.
In the panel (a), the AMSSn and ALSS no broad H$\beta$
AGNs are marked 
with large crosses. In the panel (b), 
two AGNs without significant broad line,
\ax{160118$+$0844(NO53)} and \ax{233200$+$1945(NO18)},
are marked with large crosses.
Small dots indicate the AGNs with a large optical to hard 
X-ray flux ratio ($\log f_{\rm 2-10~keV}/f_{R} > +1$).
It should be noted that the upper envelopes of the 
hydrogen column density distributions represent 
the selection limit of the 2--10~keV selection.
\label{lum_nh}}

\figcaption{
$J-K_S$ color versus redshift for 
AMSSn (open circles) and ALSS AGNs (open squares).
AGNs which are not detected in 2MASS survey are
plotted below $J-K_S$ color of 0. 
AGNs with $N_{\rm H} > 1\times10^{22}$ cm$^{-2}$ are marked
with dots. The no broad H$\beta$ AGNs
are marked with large crosses.
The small dots and crosses indicate
color distribution of optically-selected AGNs measured in 
2MASS survey (Barkhouse and Hall 2001) and Seyfert galaxies
(Kotilainen Ward and Boisson 1992), respectively. The $J-K_S$ color
track of an elliptical galaxy is shown by the solid line.
No evolution is considered in the color model.
The reddest object in the AMSSn sample is 
\ax{121854$+$2957(NO07)}.
\label{z_jk}}

\figcaption{
Luminosity distributions of $z<0.6$
AMSSn, ALSS, and ADSL AGNs (open histogram) and 
AGNs with $N_{\rm H} > 10^{22}$ cm$^{-2}$
in the sample (shaded histogram).
\label{Ldis_B}}


\figcaption{
Luminosity distributions of $z<0.6$
AMSSn, ALSS, and ADSL AGNs (open histogram) and 
no broad H$\beta$ AGNs in the sample (shaded histogram).
\label{Ldis_C}}

\figcaption{
Radio 1.4GHz luminosity versus V-band absolute magnitude
for AMSSn (open circles) and ALSS (open squares) AGNs. 
Asterisks and crosses represent upper limits for 
AMSSn and ALSS AGNs which are not detected in the 
radio wavelength, respectively. 
Small dots indicate absorbed AGNs with
$N_{\rm H} > 1\times10^{22}$ cm$^{-2}$. 
BL Lac objects are marked with large circles. 
The dashed line represents the radio-loud AGN 
criterion from Hooper et al. (1996)
\label{radio_opt1}}

\figcaption{
Radio 1.4GHz luminosity versus 
absorption-corrected 2--10~keV luminosity
for AMSSn (open circles) and ALSS (open squares) AGNs. 
Asterisks and crosses represent upper limits for 
AMSSn and ALSS AGNs which are not detected in the 
radio wavelength, respectively. 
Small dots indicate absorbed AGNs with
$N_{\rm H} > 1\times10^{22}$ cm$^{-2}$. 
BL Lac objects are marked with large circles. 
Large squares indicate AGNs which are classified as
radio-loud with radio to optical flux ratio
($\log R_{\rm 8.4GHz}>1$).
The lower and upper dashed lines indicate
typical hard X-ray to radio luminosity ratio
of radio-loud and radio-quiet QSOs, respectively 
(Elvis et al. 1994). The proposed criterion for
radio-loud AGNs is 
shown with the solid line.
\label{radio_x1}}

\newpage
\plotone{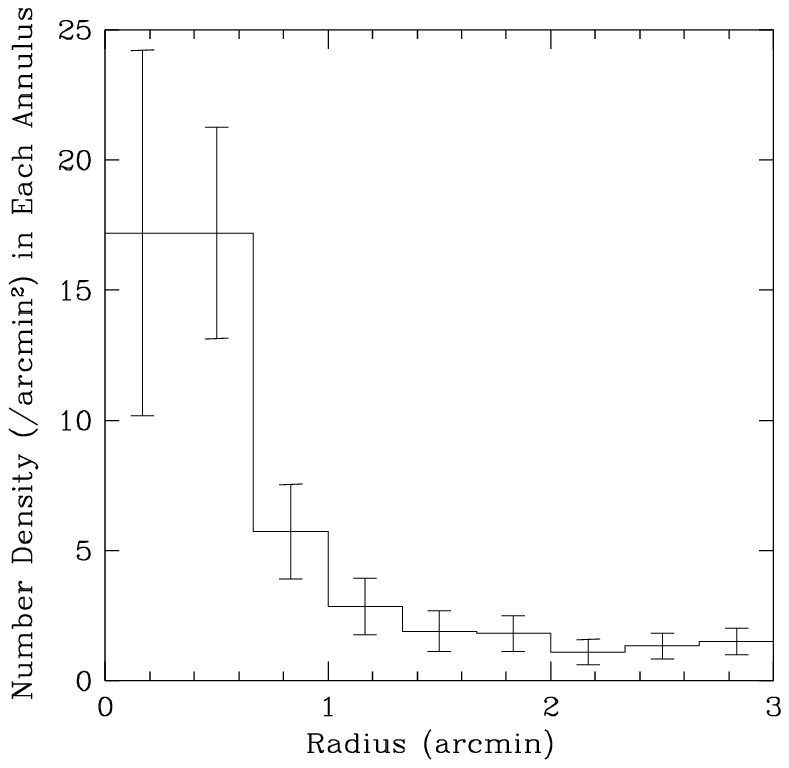}
\newpage
\plotone{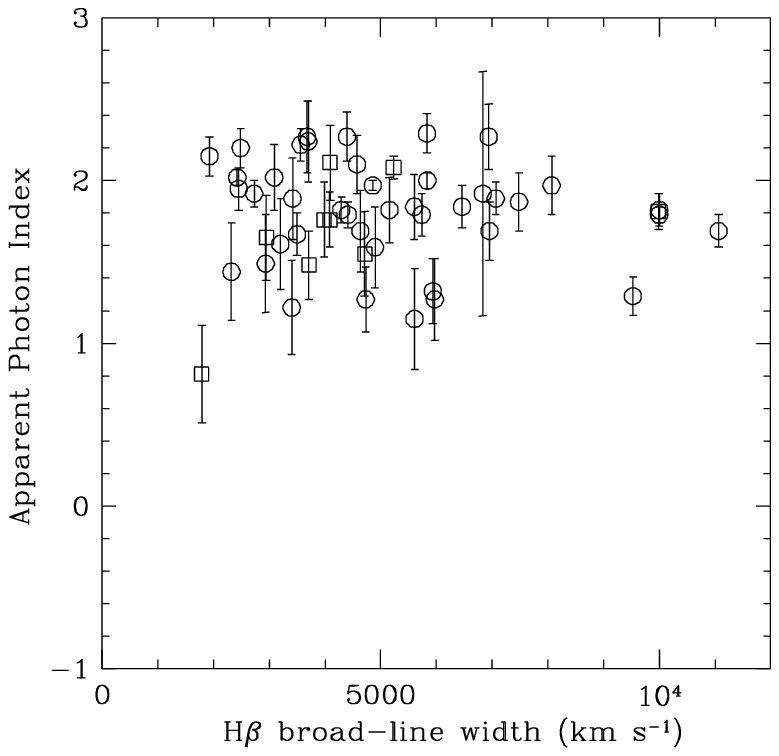}
\newpage
\plotone{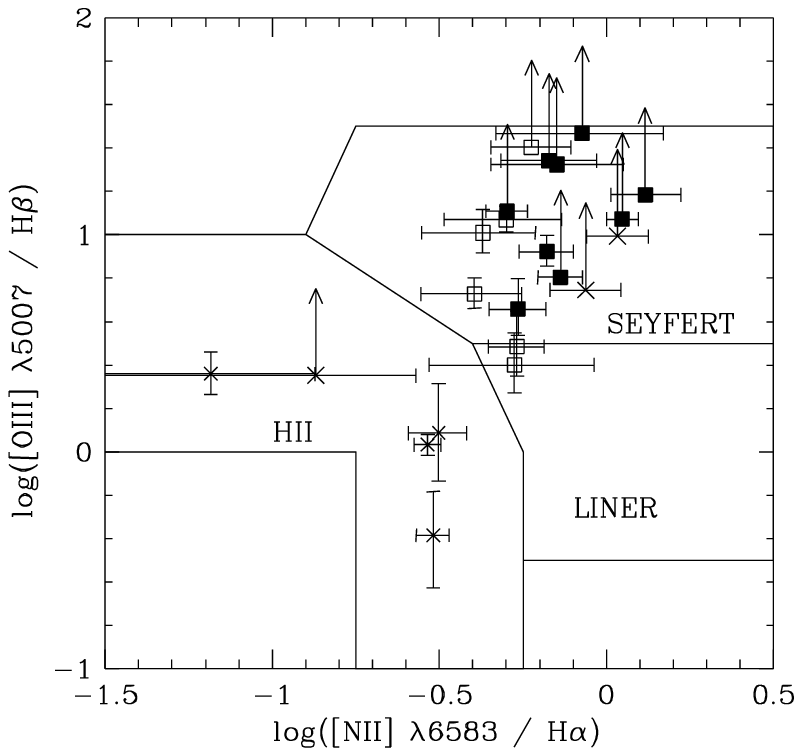}
\newpage
\plotone{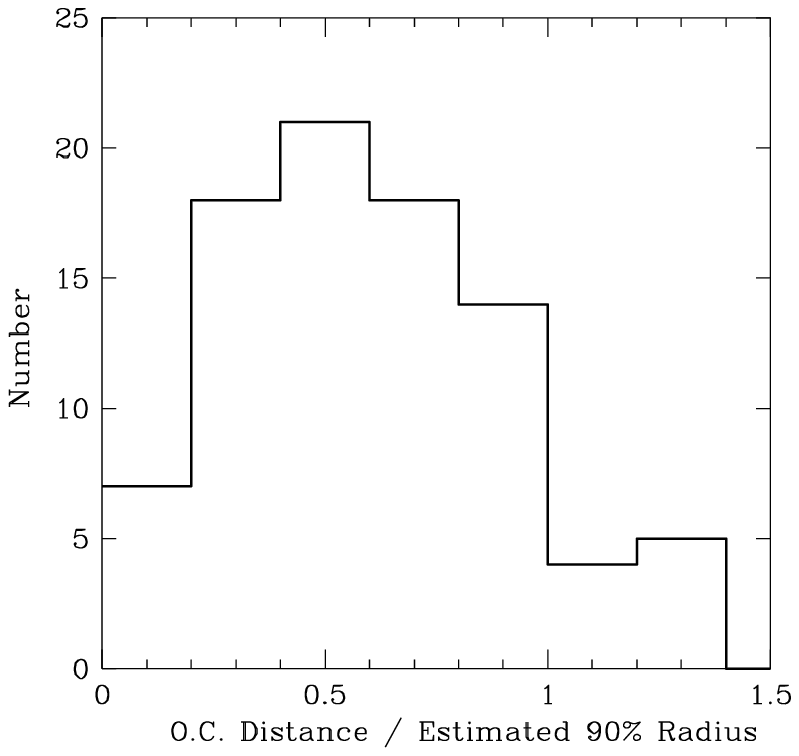}
\newpage
\plotone{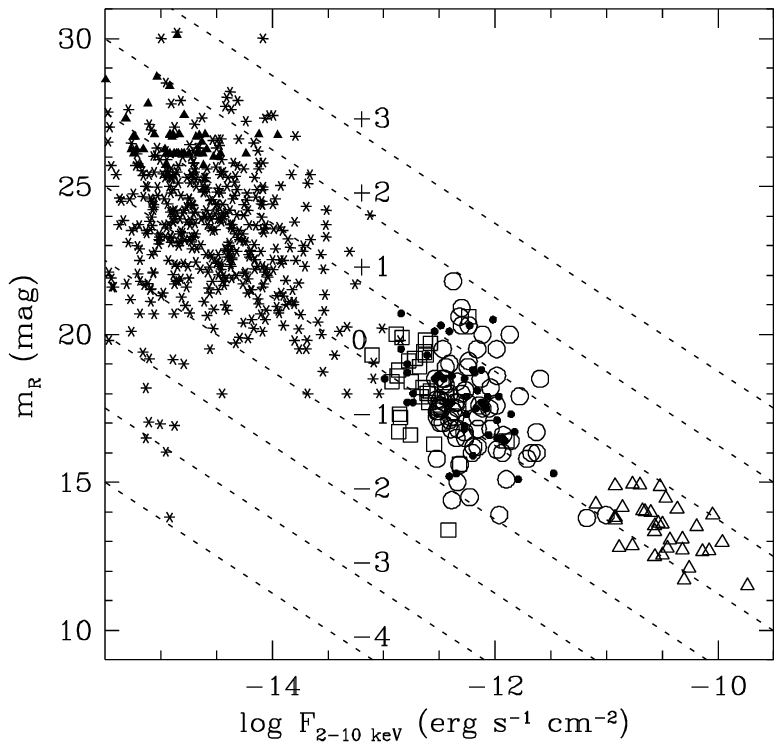}
\newpage
\plotone{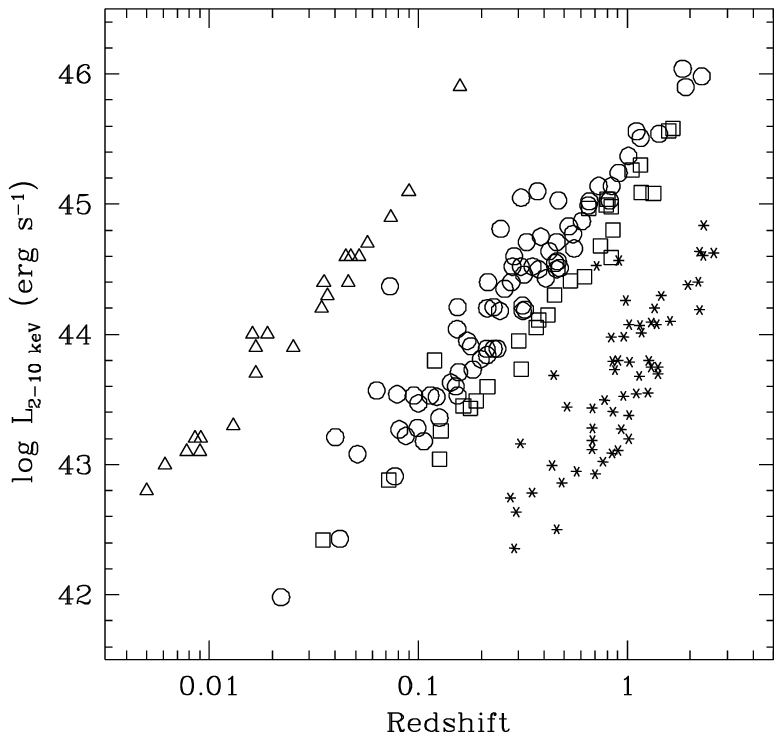}
\newpage
\plotone{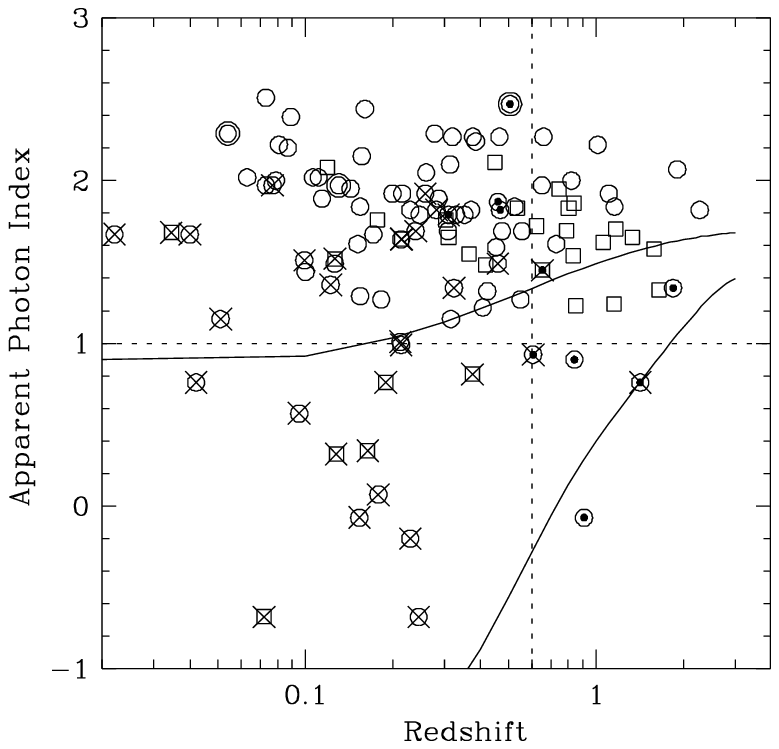}
\newpage
\plotone{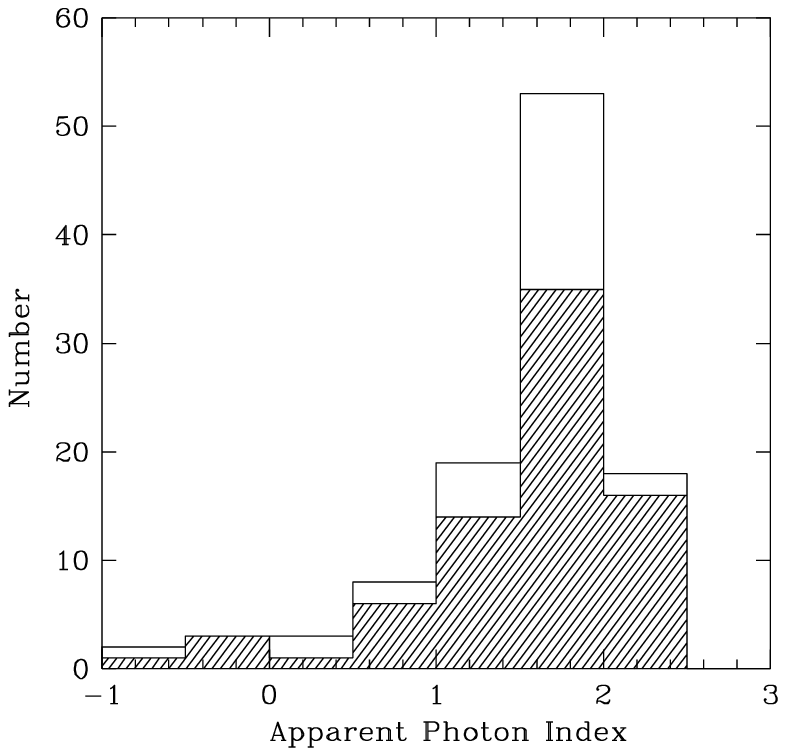}
\newpage
\epsscale{0.65}
\plotone{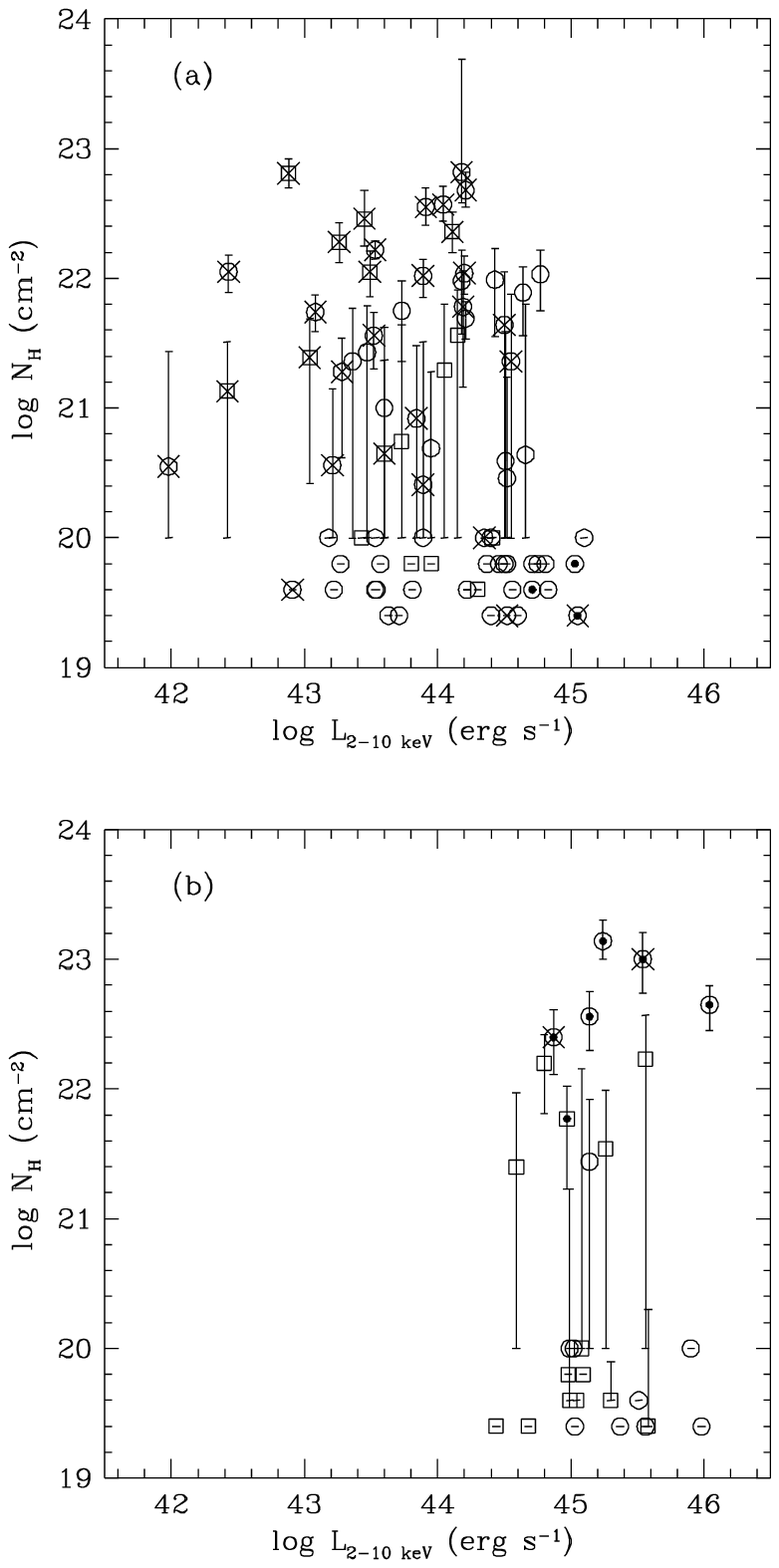}
\newpage
\epsscale{1.0}
\plotone{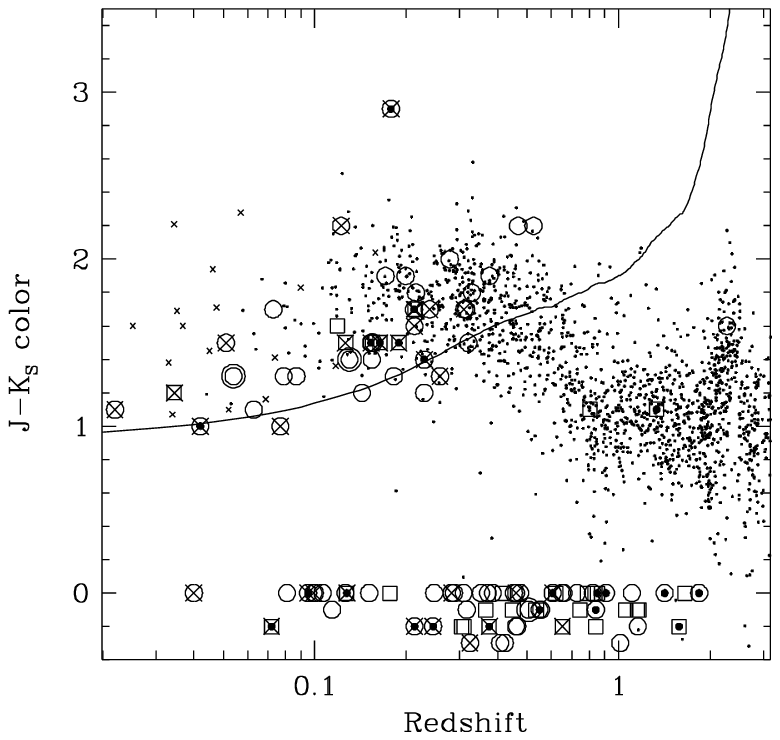}
\newpage
\plotone{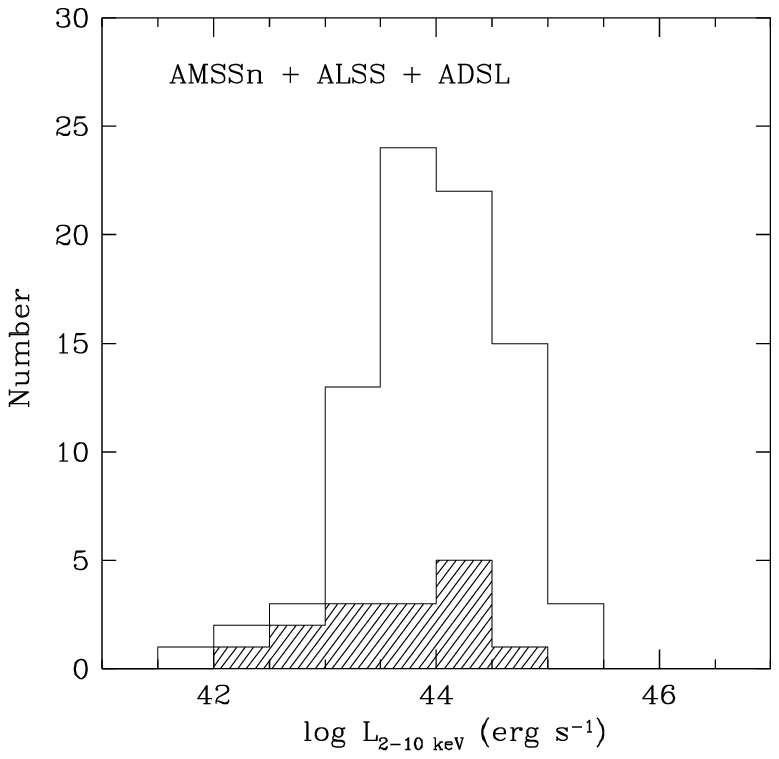}
\newpage
\plotone{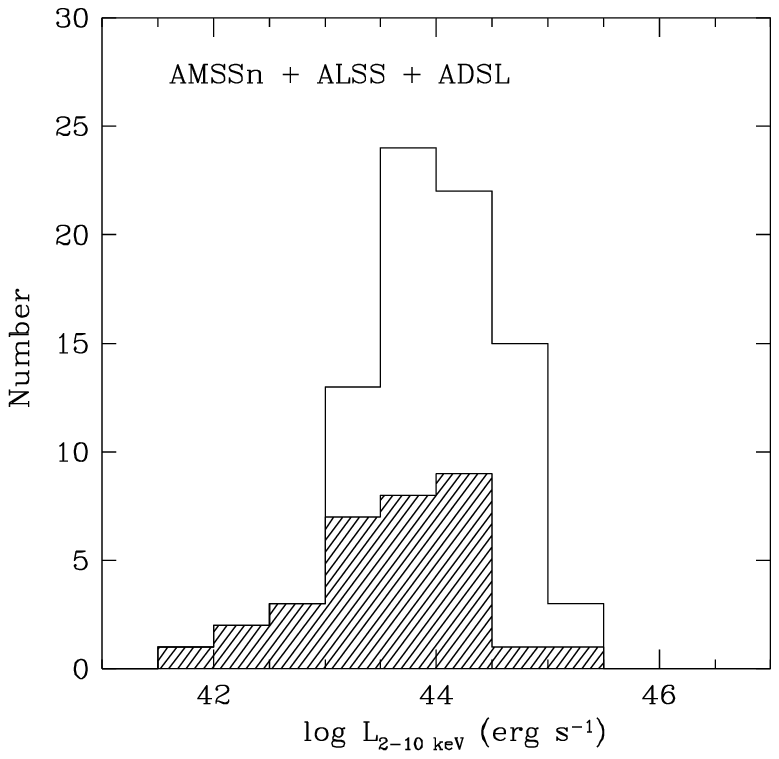}
\newpage
\plotone{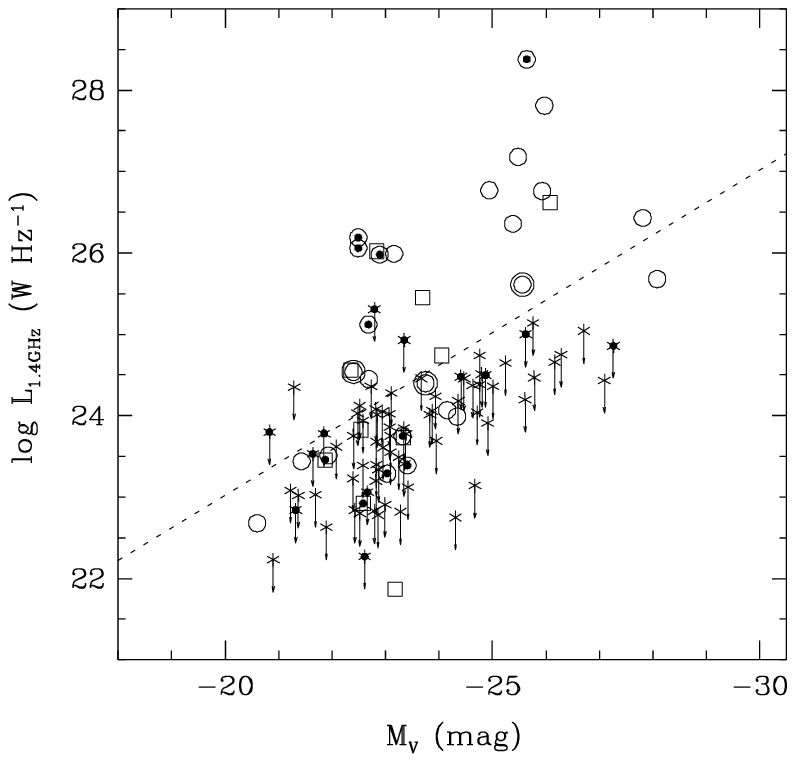}
\newpage
\epsscale{1.0}
\plotone{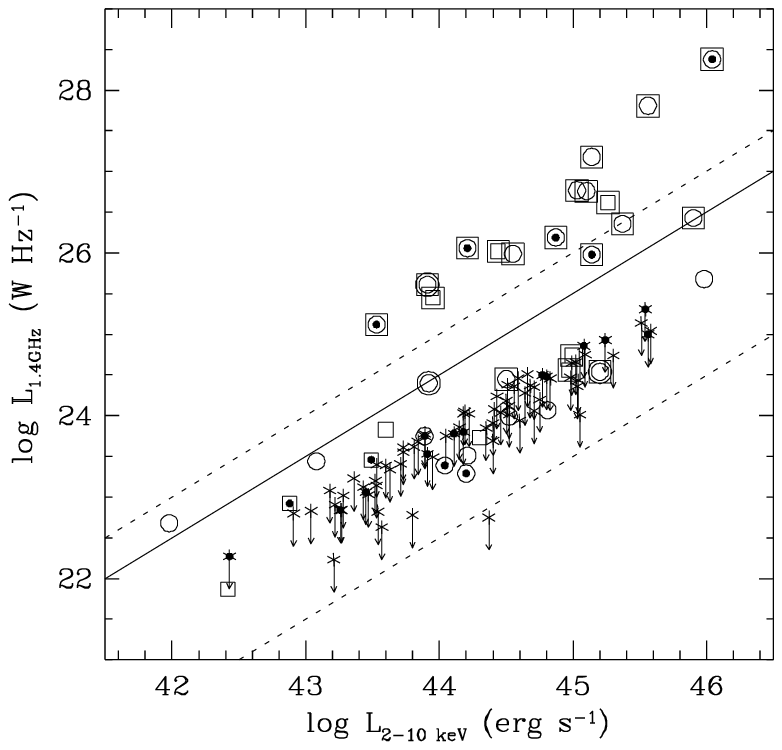}

\end{document}